\documentclass[11pt,a4paper]{article}

\usepackage[T1]{fontenc}
\usepackage[latin1]{inputenc}
\usepackage[nosort]{cite}
\usepackage{color}
\usepackage[bulletsep]{collref}
\usepackage{graphicx}
\usepackage{bbm}
\usepackage{amsmath}
\usepackage{amssymb}
\usepackage{verbatim}
\usepackage{multirow}
\usepackage{tikz}
\usepackage{amsthm}
\usepackage{fancyhdr}
\usepackage{wrapfig}
\usepackage{hyperref}
\usepackage[list=true, labelfont=bf, 
labelformat=brace, position=top]{subcaption}   
\usepackage{changes}\newcommand{\stkout}[1]{\ifmmode\text{\sout{\ensuremath{#1}}}\else\sout{#1}\fi}\setdeletedmarkup{\stkout{#1}}%for comments

\makeatletter \@addtoreset{equation}{section} \makeatother

\makeatletter
\let\old@startsection=\@startsection
\let\oldl@section=\l@section
\renewcommand{\@startsection}[6]{\old@startsection{#1}{#2}{#3}{#4}{#5}{#6\mathversion{bold}}}
\renewcommand{\l@section}[2]{\oldl@section{\mathversion{bold}#1}{#2}}
\makeatother

\makeatletter
\let\old@makecaption=\@makecaption
\def\@makecaption{\small\old@makecaption}
\makeatother

\let\oldPhi=\Phi
\let\oldPsi=\Psi
\let\oldGamma=\Gamma
\let\oldDelta=\Delta
\let\oldSigma=\Sigma
\let\oldTheta=\Theta
\let\oldPi=\Pi
\let\oldUpsilon=\Upsilon
\renewcommand{\Phi}{\mathnormal{\oldPhi}}
\renewcommand{\Psi}{\mathnormal{\oldPsi}}
\renewcommand{\Gamma}{\mathnormal{\oldGamma}}
\renewcommand{\Sigma}{\mathnormal{\oldSigma}}
\renewcommand{\Delta}{\mathnormal{\oldDelta}}
\renewcommand{\Theta}{\mathnormal{\oldTheta}}
\renewcommand{\Pi}{\mathnormal{\oldPi}}
\renewcommand{\Upsilon}{\mathnormal{\oldUpsilon}}

\newcommand{\Reals}{\mathbb{R}}

\makeatletter
\newlength{\apb@width}
\newcommand{\autoparbox}[2][c]{\settowidth{\apb@width}{#2}\parbox[#1]{\apb@width}{#2}}
\newcommand{\includegraphicsbox}[2][]{\autoparbox{\includegraphics[#1]{#2}}}
\makeatother
\ifx\genfrac\sdflkaj
\newcommand{\atopfrac}[2]{{{#1}\above0pt{#2}}}
\else
\newcommand{\atopfrac}[2]{\genfrac{}{}{0pt}{}{#1}{#2}}
\fi

\newcommand{\grp}[1]{\mathrm{#1}}

\newcommand{\vev}[1]{\langle#1\rangle}

\makeatletter
\def\mr@ignsp#1 {\ifx\:#1\@empty\else #1\expandafter\mr@ignsp\fi}%
\newcommand{\multiref}[1]{\begingroup%\let\protect\string%
\xdef\mr@no@sparg{\expandafter\mr@ignsp#1 \: }%
\def\mr@comma{}%
\@for\mr@refs:=\mr@no@sparg\do{\mr@comma\def\mr@comma{,}\ref{\mr@refs}}%
\endgroup}
\makeatother

\newcommand{\hypref}[2]{\ifx\href\asklfhas #2\else\href{#1}{#2}\fi}
\newcommand{\Secref}[1]{Section~\multiref{#1}}
\newcommand{\secref}[1]{Section~\multiref{#1}}

\newcommand{\Figref}[1]{Figure~\multiref{#1}}
\newcommand{\figref}[1]{Figure~\multiref{#1}}
\renewcommand{\eqref}[1]{(\multiref{#1})}

\usepackage{xcolor}
\ifx\href\asklfhas\newcommand{\href}[2]{#2}\fi

\newcommand{\be}{\begin{eqnarray}}
\newcommand{\ee}{\end{eqnarray}}

\begin{document}

\thispagestyle{empty}

\begin{flushright}\footnotesize
%\texttt{arXiv:xxxx.xxxx}\\
\texttt{HU-EP-18/16\\CERN-TH-2018-125}%
\end{flushright}
\vspace{2cm}

\begin{center}%
{\LARGE\textbf{\mathversion{bold}%
Consistent Conformal Extensions of the Standard Model
}\par}

\vspace{1.2cm}

\large \textsc{Florian Loebbert${}^{1}$, Julian Miczajka${}^{1}$, Jan Plefka${}^{1,2}$} \vspace{8mm} \\
\large\textit{%
${}^{1}$Institut f\"{u}r Physik, Humboldt-Universit\"{a}t zu Berlin, \\
Zum Gro{\ss}en Windkanal 6, 12489 Berlin, Germany
} \\[0.3cm]
 \textit{${}^{2}$Theoretical Physics Department, CERN\\ 1211 Geneva 23, Switzerland}

\texttt{\\ \{loebbert,miczajka,plefka\}@physik.hu-berlin.de}

%%%%%%%%
\par\vspace{15mm}

\textbf{Abstract} \vspace{5mm}

\begin{minipage}{12.2cm}
The question of whether classically conformal modifications of the standard model are consistent with experimental obervations has recently been subject to renewed interest. The method of Gildener and Weinberg provides a natural framework for the study of the effective potential of the resulting multi-scalar standard model extensions. This approach relies on the assumption of the ordinary loop hierarchy $\lambda_\text{s} \sim g^2_\text{g}$ of scalar and gauge couplings. On the other hand, Andreassen, Frost and Schwartz recently argued that in the (single-scalar) standard model, gauge invariant results require the consistent scaling $\lambda_\text{s} \sim g^4_\text{g}$. In the present paper we contrast these two hierarchy assumptions and illustrate the differences in the phenomenological predictions of minimal conformal extensions of the standard model.
\end{minipage}
\end{center}

%%%%%%%%%%%%%%%%%%%%%%%%%%%%%%%%%%%%%%%%%%%%%%%%%%%%%%%%%%%%%%%%%%%%%%%%%%%
%%%%%%%%%%%%%%%%%%%%%%%%%%%%%%%%%%%%%%%%%%%%%%%%%%%%%%%%%%%%%%%%%%%%%%%%%%%
\newpage

\tableofcontents

\bigskip
\noindent\hrulefill
\bigskip

%%%%%%%%%%%%%%%%%%%%%%%%%%%%%%%%%%%%%%%%%%

%%%%%%%%%%%%%%%%%%%%%%%%%%%%%%%%%%%%%%%%%%%%%%%%%%%%%%%%%%%%%%%%%%%%%%%%%%%

\section{Introduction}
\label{sec:intro}

The standard model (SM) of particle
physics represents a quantum field theory staying perturbatively consistent under renormalization group (RG) flow all the way up to the Planck scale $M_{\text{Pl}}$, where 
quantum gravity effects become relevant.
Due to the absence of any beyond-the-standard-model signals from the LHC, the conservative
scenario of `no (or minimal) new physics up to $M_{\text{Pl}}$' has gained some momentum in the
community. Still, there
are a number of obvious shortcomings. For one, the Higgs mass parameter is unnaturally
small compared to $M_{\text{Pl}}$, known as the hierachy problem. Moreover, the strong observational evidence for dark matter and neutrino masses calls for an extension of the SM. Ideally, such an extension should also resolve 
the puzzle of the metastable Higgs vacuum \cite{Degrassi:2012ry,Buttazzo:2013uya} .

A good guideline for extending the SM is to call for additional symmetries, such as supersymmetry, which should be broken in order to agree with current experimental observations. In this paper we focus on another prominent example of a guiding (and broken) symmetry: The standard model is `nearly' conformal as a classical field theory. Conformal invariance is only broken by the explicit Higgs field mass term, which induces the electroweak symmetry breaking and the masses of all known elementary particles. At least qualitatively, the same effect can be generated in a
classically scale-free model, whose radiative corrections lead to a spontaneous generation of mass scale and symmetry breaking as was first advocated by Coleman and E.~Weinberg \cite{Coleman:1973jx} (see \cite{Sher:1988mj} for an extensive review). 
While classical conformal symmetry is broken via radiative corrections, the 
vanishing mass term is stable under renormalization \cite{Bardeen:1995kv} using
dimensional regularization. However, this attractive scenario 
does not yield a radiative breaking
of the electroweak symmetry in a scalefree version of the SM (simply dropping the scalar
mass term) due to the largeness of the top mass. By adding additional
bosonic degrees of freedom to the SM --- e.g.\ via an extended Higgs sector or novel gauge
fields --- this problem may in principle be cured. 
The question for realizing such a (minimal) version of a conformal standard model has recently attracted a
lot of attention in the literature, see e.g.\ \cite{Fatelo:1994qf,Hambye:1995fr,Hempfling:1996ht,Meissner:2006zh, Chang:2007ki,Foot:2007as, Foot:2007iy,Iso:2009ss,Iso:2009nw, AlexanderNunneley:2010nw,Carone:2013wla,Englert:2013gz,Farzinnia:2013pga,Hambye:2013sna,Heikinheimo:2013fta,Holthausen:2013ota,Hill:2014mqa, Karam:2015jta, Oda:2015gna,Das:2016zue,Helmboldt:2016mpi, Hambye:2018qjv}. A common feature of these works is an extended scalar
sector for which an effective potential needs to be established, minimized and
spontaenously generated masses extracted in a perturbatively
consistent and also gauge invariant fashion. As we shall argue in this paper this is
a non-trivial issue building upon certain scaling assumptions of the scalar couplings
$\lambda_{i}$ with respect to the gauge and Yukawa couplings.

%%%%%%%%%%%%%
Besides the Coleman--Weinberg mechanism, another way to break conformal symmetry is via introduction of a (typically high) cutoff scale $\Lambda$. This scale induces quadratic divergences generating a Higgs mass contribution proportional to $\Lambda^2$. In the context of conformal extensions of the standard model,
this naturalness problem is an often addressed point of criticism (see e.g.\ \cite{Antipin:2013exa,Chankowski:2014fva}), since the classically vanishing Higgs mass requires fine tuning in order to stay small.
Here we do not employ a cutoff scale. We perform all calculations using dimensional regularization and employ
 the $\overline{\text{MS}}$-renormalization scheme. We will also not address the question of embedding the considered models into a theory of quantum gravity which would result in a natural cutoff $\Lambda\sim M_\text{Pl}$ at the Planck scale.

\paragraph{Multi-scale Issue.}
An immediate problem that one faces when considering multi-scalar conformal extensions of the standard model is the question of multi-scale renormalization.
Loop contributions to the effective potential typically come with logarithms that depend on the ratio of the scalar field and the renormalization scale. These contributions become large when the field value and the scale differ significantly, which invalidates the perturbative expansion. In the case of a single scalar field, a single renormalization scale is sufficient to resum these logarithmic contributions via the renormalization group (RG) which results in the RG-improved effective potential. What is done here is to set the arbitrary
renormalization scale $\mu$ to the scalar field value at its minimum $\vev{\phi}$ which
cancels all logarithms $\log \frac{\phi}{\vev{\phi}}$ in the potential and derived quantities
thereof in the vacuum configuration.
However, in the presence of multiple scalar fields one faces logarithms of different field-to-scale ratios, which renders a single renormalization scale insufficient for dealing with all logarithmic contributions at the same time.
 A natural resolution seems to be the introduction of multiple renormalization scales as proposed by Einhorn and Jones \cite{Einhorn:1983fc} and later refined by Ford and Wiesendanger \cite{Ford:1994dt,Ford:1996hd,Ford:1996yc}. At least in certain cases, this approach was argued \cite{Casas:1998cf} to be equivalent to the decoupling method of \cite{Bando:1992wy}, which splits the mutli-scale problem into single-scale problems in between different mass thresholds. Unfortunately, these methods complicate the RG-analysis significantly which limits their applicability to simpler toy models or low loop orders. A more powerful approach was suggested by Gildener and S.~Weinberg \cite{Gildener:1976ih}: Assuming the presence of a flat direction of the classical potential in the space of scalar fields at some renormalization
 scale $\mu_\text{GW}$  the symmetry breaking is studied only in this direction, which again results in a single-scale problem. In the context of conformal extensions of the standard model, this approach was recently applied in e.g.~\cite{Helmboldt:2016mpi}. A new method for the study of multi-scale potentials was suggested in \cite{Chataignier:2018aud}, which assumes the existence of a (field-dependent) value of the renormalization scale where all loop corrections to the effective potential vanish. Working at this scale translates the problem of understanding the full effective potential into a study of the tree-level potential with running coupling constants.

Notably, even in a (single-scale) textbook approach the problem of multi-scale renormalization can be avoided in certain theories at one-loop order. This is due to a special prescription for solving the minimum conditions for the effective potential as illustrated in \Secref{sec:QP} for the model of \cite{Hempfling:1996ht}.

\paragraph{Gauge Dependence Issue.}
Besides the problem of multi-scale renormalization, another important requirement is gauge independence.
The effective potential is generically gauge dependent \cite{Jackiw:1974cv,Dolan:1974gu} and great care is needed to extract physical information contained in its minimal value. 
However, drawing conclusions on which modifications of the standard model are compatible with current experimental data is very sensitive to small modifications and thus it requires caution to identify a minimal model. In particular, the gauge dependence of the effective potential has recently been emphasized in \cite{Andreassen:2014eha,Andreassen:2014gha}. In the context of the standard model, gauge invariance was shown to require a non-standard hierarchy of coupling constants \cite{Andreassen:2014gha}, which amounts to taking the scalar coupling(s) $\lambda_{s}$ to be of the order
of the fourth power of the gauge couplings $g_{i}$ and Yukawa couplings $y_{t}$, i.e.\
\begin{equation}\label{eq:hierarchy1}
\lambda_{s} \sim \mathcal{O}(g_{i}^{4})  \sim \mathcal{O}(y_{t}^{4}).
\end{equation} 
This hierarchy has also been shown to hold in the Coleman--Weinberg model \cite{Coleman:1973jx}. Reintroducing $\hbar$ this scaling is easily motivated by setting $\lambda_{s}\sim
\hbar$, in other words the classical scalar potential is made quantum by hand. Although being
rather unconventional, this choice clearly allows the leading scalar potential to receive
seizable one-loop quantum corrections that can significantly shift its minimum to non-vanishing field values $\vev{\phi}\neq 0$. As this scaling assumption makes the (tree-level) scalar
potential essentially quantum, we term it the `Quantum Potential' approach.

The ordinary loop counting on the other hand amounts to the assumption
\begin{equation}\label{eq:hierarchy2}
\lambda_{s} \sim \mathcal{O}(g_{i}^{2})  \sim \mathcal{O}(y_{t}^{2})\, ,
\end{equation}
which is equivalent to taking $\lambda_{s}$ not to be of order $\hbar$ as is usually done. 
In fact this is the scaling applied in the Gildener--Weinberg scheme.

\paragraph{The Setup.}
Let us explain the difference of the generic situation we are facing in multi-scalar
extensions of the conformal SM in more detail.
Due to the assumed classical scale invariance the tree-level part of the scalar potential needs
to be of the form
\be
V_{0}(\vec\Phi)=\frac{1}{4}\lambda_{IJKL}\Phi_{I}\Phi_{J}\Phi_{K}\Phi_{L},
\ee
where the totally symmetric symbol $\lambda_{IJKL}$ parametrizes the set  of scalar
couplings and the $\Phi_{I}(x)$
denote all scalar fields in the theory. The one-loop correction to the scalar potential
computed in dimensional regularization then takes the generic form
\be
V_{\text{eff}}^{(1)}=\frac{\hbar}{64\pi^{2}}\sum_{i} n_{i}\, m_{i}[\Phi_{I}]^{4}
\left (  \ln \frac{m_{i}[\Phi_{i}]^{2}}{\mu^{2}} - a_{i}-\frac{1}{\epsilon}\right ).
\ee
Here the index $i$  runs over all particles in the theory which couple
to the scalars. For each particle, $m_{i}[\Phi_{I}]$ denotes its field-dependent effective
\emph{tree-level} mass, which emerges for non-zero scalar vacuum expectation values and
 implicitly depends on the renormalization scale $\mu$.
The $n_{i}$ count the real degrees of freedom of the particle $i$ with a minus sign
for fermions, while the $a_{i}$ are scheme dependent constants: in the $\overline{\text{MS}}$
scheme they are given by $-5/6$ for gauge bosons and $-3/2$ for fermions or scalars.
Clearly, the classical potential always has the trivial vacuum $\vev{\vec{\Phi}}=0$.
The quest is now to have $V=V_{0}+V_{\text{eff}}^{(1)}$ develop a minimum at a
nonzero value of $\vev{\vec{\Phi}}$ through radiative corrections. Here we focus on the comparison of two scenarios:
\begin{enumerate}
\item {\bf Quantum-Potential Approach:} 
The multi-scalar effective potential is treated as in the single-scalar case, e.g.\ in the case of the standard model.
The scalar couplings $\lambda_{IJKL}$ are taken to scale as $\hbar$, i.e.\ to be of
the same order of magnitude as (part of) the one-loop contributions $V_{\text{eff}}^{(1)}$ to the effective potential. Effectively, this amounts to assuming a hierarchy of couplings $\lambda\sim g^{4}$,
where $g$ are the gauge or Yukawa couplings. This scaling hierachy pushes
the scalar coupling contribtions in $V^{(1)}_{\text{eff}}$ to the next order.
\item {\bf Gildener--Weinberg Method:} One takes $V_{0}$ to have a degenerate zero energy 
vacuum along a ray $\vev{\Phi}=\varphi\, \vec{n}$, parametrized by a sliding-scale field $\varphi$ \emph{at a particular scale} $\mu_\text{GW}$.
The quantum fluctuations of $V_{\text{eff}}^{(1)}$ then lift the degeneracy along this valley
and yield a radiatively generated non-vanishing vacuum expectation value $\vev{\vec{\Phi}}$, which induces all
 the masses in the theory. In this approach, the ordinary loop hierarchy of couplings $\lambda\sim g^2$ is assumed. Importantly, finding the minimum of the multi-scalar effective potential reduces to a single scalar problem for the field $\varphi$.
\end{enumerate}

In the case of single scale models, it can be seen that both scenarios yield the same result at one-loop order.
In the present paper we contrast these two different methods in the multi-scalar case (which correspond to the two hierarchy assumptions given above) in order to extract gauge invariant data from the effective potential.

In the following analyses we neglect influences of all leptons except for the top quark. Furthermore, since we work entirely at one-loop level and the Higgs boson has no color charge, we can neglect all contributions to the effective potential that come from the strong interaction. They will be of higher loop order.

The paper is structured as follows.
We start by introducing the Hempfling model of \cite{Hempfling:1996ht} in \secref{sec:Hempfling} as our laboratory throughout the paper. We proceed to apply the Quantum Potential (QP) approach and the Gildener--Weinberg (GW) method in \secref{sec:QP} and \secref{sec:GW}, respectively. In particular, for both cases we determine the allowed mass ranges for the new scalar and the new $\grp{U}(1)$ gauge boson $Z'$, as well as the allowed couplings, demanding compatibility with experimental bounds on the scalar mixing and the absence of Landau poles and vacuum instability up to the Planck scale. In \secref{sec:specialcase} we briefly demonstrate that a further reduction of the field content does not lead to a phenomenologically viable model. Finally we conclude by comparing the QP and GW approaches and give a brief outlook.

\paragraph{Note added:}
On the day of posting this manuscript the article \cite{Chataignier:2018kay} appeared on the arXiv which has some overlaps with the ideas of our article and contains interesting complementary results.

%%%%%%%%%%%%%%%%%%
\section{Our Laboratory: The Hempfling Model}
\label{sec:Hempfling}

It is well known that implementing Coleman--Weinberg symmetry breaking into the standard model with vanishing mass term does not give rise to a phenomenologically viable vacuum due to the large top mass. As we will see below the same applies to an extension of this model by a single scalar field, see also \cite{Helmboldt:2016mpi}. The addition of new fermionic degrees of freedom will give negative contributions to the mass eigenvalues and only worsen the situation. We are thus led to introduce new bosonic fields. Restricting to renormalizable models we can add scalar fields or vector fields. 

\paragraph{Hempfling Model.}
Here we will analyse the conformal extension of the standard model that was proposed by Hempfling already in 1996, i.e.\ before the discovery of the Higgs boson. In addition to the standard model at vanishing tree-level Higgs mass, the Hempfling model contains  a new `dark' 
$\grp{U}(1)$ gauge boson $Z_{\mu}'$ exclusively coupled to a new scalar field $S$. This is a minimal extension of the conformal standard model in the following sense: As we will illustrate in section 5 an extension by a single scalar field $S$ alone is not capable of consistently reproducing the correct Higgs mass. Hence, we must add additional bosonic degrees of freedom. Adding yet another scalar would introduce several new couplings to the Higgs and the scalar $S$. We thus add a new abelian gauge field $Z'_\mu$ coupled only to the new scalar $S$ by the new gauge coupling $g_{Z'}$.
The complex scalar $S$ has a $\grp{U}(1)$ phase symmetry
$S \rightarrow e^{i\alpha} S$,
and the full Lagrangian for this model is given by 
\begin{equation}
\mathcal{L}_{\text{Hempf}} =\  \mathcal{L}_\text{SM}\big|_{\atopfrac{\lambda=0}{m_\text{H}=0}}- V(H, S)+ \mathcal{D}_\mu S (\mathcal{D}^\mu  S)^{\dagger} -\frac{1}{4}F'_{\mu\nu}{F'}^{\mu\nu} 
+ \mathcal{L}_{\text{GF}} + \mathcal{L}_{\text{ghosts}},
\end{equation}
where $ \mathcal{L}_\text{SM}\big|_{\atopfrac{\lambda=0}{m_\text{H}=0}}$ is the SM
Lagrangian without the Higgs potential. The new tree-level potential is given by
\begin{equation}
V(H,S) = \  \lambda_1 (H^\dagger H)^2 + \lambda_{12} (H^\dagger H)(S^\dagger S) + \lambda_2 (S^\dagger S)^2.
\end{equation}
We work with the gauge fixing terms
\begin{equation}
\mathcal{L}_{\text{GF}} = -\frac{1}{2\xi_{B}}(\partial_{\mu}B^{\mu})^{2}
-\frac{1}{2\xi_{W}}(\partial_{\mu}A^{a\, \mu})^{2}-\frac{1}{2\xi_{Z'}}(\partial_{\mu}
Z^{'\mu})^{2}\, ,
\end{equation}
keeping the $\xi_{i}$ arbitrary.
Note that we also do not consider $U(1)$ mixings $F_{\mu\nu}{F'}^{\mu\nu}$ with the photon $A_{\mu}$.
Next the Higgs doublet and the new scalar are written in a background field $(\hat\phi,
\hat S)\in\Reals$ expansion as
\begin{align}
H &= \frac{1}{\sqrt{2}} \begin{pmatrix} \phi_1 + i \psi_1 \\  \hat \phi + \phi_2 + i \psi_2\end{pmatrix},
&S = \frac{1}{\sqrt{2}}( \hat  S + s_{1}+ i s_{2}). \label{expansion}
\end{align}
Terms involving ghosts or the strong interactions may be left out, as they only appear at higher loops.
 The covariant derivative couples $S$ to the new gauge field $Z'^\mu$ according to
\begin{align}
\mathcal{D}_\mu S = \left(\partial_\mu + i g_{Z'} Z'_\mu\right)S,
\end{align}
inducing cubic and quartic interactions. Because the new gauge field does not interact directly with the fields of the standard model, we call it a `dark' gauge field.

%%%%%%%%%%%%%%%%%%%%%%%%%%%%%%%%%%%%%%%%%%%%%%%%%%%%%%%%%%%%%%%%%%%%%

\paragraph{Effective Potential.}
As usual we expand the scalar fields around classical field values $(\hat\phi,
\hat S)$ and integrate out the quantum fields to arrive at the effective potential at one-loop order. Dropping from now
on the hats of the background fields we have
\begin{align}
V_\text{eff}(\phi, S) = \frac{\lambda_1}{4}\phi^4 + \frac{\lambda_{12}}{4}\phi^2 S^2 + \frac{\lambda_2}{4} S^4 + \frac{1}{64 \pi^2}\sum_{i\in I} n_i m_i^4 \left(\ln \frac{m_i^2}{\mu^2} - a_i -\frac{1}{\epsilon}\right),
\label{54}
\end{align}
with $I=\{A, B, C, E_{\pm}, F_{\pm}, G_{\pm}, I_{\pm}, T\}$. The field dependent masses are given by
\begin{align}
m_A^2 &= \frac{g_2^2}{4} \phi^2, \qquad m_B^2 = \frac{\left(g_1^2+g_2^2\right)}{4} \phi^2, \qquad m_C^2 =  g_{Z'}^2 S^2, \qquad m_T^2 = \frac{y_t^2}{2} \phi^2,
\end{align}
as well as
\begin{align}
m_{E\pm}^2 &= \frac{1}{4} \left(\left(6 \lambda_1 + \lambda_{12}\right)\phi^2 +\left(6 \lambda_2 + \lambda_{12}\right)S^2 \pm \sqrt{\left(\left(6 \lambda_1 - \lambda_{12}\right)\phi^2 -\left(6 \lambda_2 - \lambda_{12}\right)S^2 \right)^2+16 \lambda_{12}^2\phi^2 S^2}\right)\notag\\
m_{F\pm}^2 &= \frac{1}{4}\left(2\lambda_2 S^2 + \lambda_{12} \phi^2 \pm \sqrt{\left(2 \lambda_2 S^2 + \lambda_{12} \phi^2\right)^2 -\xi_{Z'} \left(4 \lambda_2 g_{Z'}^2 S^4+ 2 \lambda_{12} g_{Z'}^2 \phi^2 S^2\right)}\right),\notag\\
m_{G\pm}^2 &= \frac{1}{4} \left(2 \lambda_1 \phi^2 + \lambda_{12}S^2 \pm \sqrt{\left(2\lambda_1 \phi^2 +\lambda_{12}S^2\right)^2 - \left(\xi_B g_1^2 + \xi_W g_2^2\right)\left(4\lambda_1 \phi^4 + 2 \lambda_{12} \phi^2 S^2\right)}\right),\notag\\
m_{I\pm}^2 &= \frac{1}{4} \left(2 \lambda_1 \phi^2 + \lambda_{12}S^2 \pm \sqrt{\left(2\lambda_1 \phi^2 +\lambda_{12}S^2\right)^2 - \xi_W g_2^2\left(4\lambda_1 \phi^4 + 2 \lambda_{12} \phi^2 S^2\right)}\right),
\label{55}
\end{align}
while the the parameters $n_i$ and $a_{i}$ take the form
\begin{align}
n_A &= 6, \quad n_B = 3, \quad n_C = 3,\quad n_T = -12,
&
n_E &= n_F = n_G = 1 = \frac{1}{2} n_I, \notag\\
a_A &= a_B = a_C = -\frac{5}{6},
&
a_E&=a_F=a_G = a_I = a_T = -\frac{3}{2}.
\end{align}
Note that of all leptons only the top quark is included in the analysis, as its couplings
$y_{t}$ is by far dominant.

%%%%
\paragraph{Beta Functions.}

The one-loop beta functions for the Hempfling model are given by \cite{Helmboldt:2016mpi, Srednicki}
\begin{align}
\label{eq:betafcts}
\beta_{\lambda_1}&=24 \lambda_1^2 + \lambda_{12}^2 - 3 \lambda_1 (g_1^2 + 3 g_2^2) + \frac{3}{8} \left(g_1^4 + 2 g_1^2 g_2^2 + 3 g_2^4\right) + 12 \lambda_1 y_t^2 - 6 y_t^4, \nonumber \\
\beta_{\lambda_{12}} &= \lambda_{12}\left(12 \lambda_1 + 8 \lambda_2 + 4 \lambda_{12} + 6 y_t^2 - \frac{3}{2} \left(g_1^2 + 3 g_2^2\right) - 6 g_{Z'}^2\right), \nonumber \\
\beta_{\lambda_2} &=20 \lambda_2^2 + 2 \lambda_{12}^2 - 6 \lambda_2 g_{Z'}^2 + 6 g_{Z'}^4 \nonumber\\
\beta_{g_1} &= \frac{41}{6} g_1^3, \qquad \beta_{g_2} = -\frac{19}{6} g_2^3, \qquad \beta_{g_{Z'}} = \frac{1}{3} g_{Z'}^3,\nonumber\\
\beta_{g_3} &= -7 g_3^3, \qquad \beta_{y_t} = y_t (\frac{9}{2} y_t^2 - \frac{17}{12} g_1^2 - \frac{9}{4} g_2^2 - 8 g_3^2), 
\end{align}
where $\beta_\alpha = 16 \pi^2 \mu \frac{\text{d}\alpha}{\text{d} \mu}$.

%%%%%%%%%%%%%%%%%

\section{Quantum Potential Approach}
\label{sec:QP}

In this section we consider the Quantum Potential method and apply it to the concrete example of the Hempfling model. We discuss the phenomenological consistency with the observed value of the Higgs mass and study the absence of Landau poles and vacuum stability up to the Planck scale.
%%%%%%%%%%%%%%%%%

\paragraph{Conceptual Idea.}

As motivated in the introduction, we will now impose the hierarchy of coupling constants 
\begin{equation}\label{eq:hierarchy}
\lambda_j\sim g_k^4\sim y_t^4\sim\hbar,
\end{equation}
which allows to consistently expand all quantities in the small parameter $\hbar$. 
This assumption extrapolates the hierarchy $\lambda\sim g^4$ of scalar electrodynamics which --- in that model ---  was explicitly proven by Coleman and E.~Weinberg using the renormalization group \cite{Coleman:1973jx}. 
Indeed, for theories with a single scalar field, this hierarchy of couplings is necessary in order for a one-loop contribution to push the minimum of the 
tree-level potential from zero field values to a non-zero value, i.e.\ to implement Coleman--Weinberg symmetry breaking. Moreover, this hierarchy is crucial to guarantee the gauge independence of physical information extracted from the Coleman--Weinberg potential \cite{Andreassen:2014eha}. 
Also for the standard model, imposing the hierarchy \eqref{eq:hierarchy} consistently guarantees trustable and gauge independent results \cite{Andreassen:2014gha}. 

 %%%%%%%%%%%%%%%%%%%%%%%%%%%%%%%%%%
 
\paragraph{QP of Hempfling Model.}
Solving the stationarity conditions
\begin{align}
0 =\left. \frac{\text{d}V_\text{eff}}{\text{d}\phi}\right|_{\phi=\langle \phi\rangle, S=\langle S \rangle}\, , \qquad
0 =\left. \frac{\text{d}V_\text{eff}}{\text{d}S}\right|_{\phi=\langle \phi\rangle, S=\langle S \rangle},
\label{eq:statcondQP}
\end{align}
for $\lambda_1$ and $\lambda_2$, we arrive at the one-loop effective potential
\begin{align}
V_\text{eff}^\text{ren}(\phi, S) = &-\frac{\lambda_{12}}{8} \vev{\phi}^2 \vev{S}^2 \left(\frac{\phi^2}{\vev{\phi}^2} - \frac{S^2}{\vev{S}^2}\right)^2 + \frac{3}{64 \pi^2} g_{Z'}^4 S^4 \left(\log \frac{S^2}{\vev{S}^2} - \frac{1}{2}\right)\nonumber\\
& + \frac{3}{64\pi^2}\frac{\phi^4}{\vev{\phi}^4}\left(2 m_W^4 + m_Z^4 - 4 m_t^4\right) \left(\log \frac{\phi^2}{\vev{\phi}^2} - \frac{1}{2}\right),
\label{eq:effpotQP}
\end{align}
where we have
\begin{align}
m_W&=\frac{g_2}{2}\vev{\phi},
&
m_Z&=\frac{\sqrt{g_1^2+g_2^2}}{2}\vev{\phi},
&
m_t&=\frac{y_t}{\sqrt{2}}\vev{\phi}.
\end{align}
Crucially, all dependence on the gauge parameters disappears from the effective potential \eqref{eq:effpotQP} after imposing the hierarchy of couplings. That is, as in the standard model, the hierarchy \eqref{eq:hierarchy} resolves the gauge dependence issue.
Moreover, solving \eqref{eq:statcondQP} symmetrically for $\lambda_1$ and $\lambda_2$ (instead of e.g.\ for $\lambda_1$ and $\lambda_{12}$), the dependence of the logarithms on the renormalization scale $\mu$ completely drops out of the effective potential. In particular, we have no multi-scale problem which naively could have been expected, cf.\ the introductory \secref{sec:intro}. This simplifies the analysis considerably. This property is a consequence of the
vanishing of $\beta_{\lambda_{12}}$ at one-loop under the scaling assumptions 
$\lambda_{i}\sim\mathcal{O}(g_{i}^{4})$ compare (\ref{eq:betafcts}).
 We do not expect it to prevail at higher loop orders.

The above effective potential has a minimum at $\{\vev{\phi},  \vev{S}\}$. Since the Higgs
vacuum expectation value is fixed by experiment, the potential contains three free parameters, namely $\{\lambda_{12}, g_{Z'}, \vev{S}\}$. However, we can fix an additional one by demanding the existence of a mass eigenstate with eigenvalue $m_h$, the measured Higgs mass.

In oder to do so, we calculate the Hessian of the potential at its
minimum 
\begin{equation}
\label{M2QP}
M_{ij}^{2}=  \frac{\partial^{2} V_\text{eff}^\text{ren}}{\partial \phi_{i}\partial \phi_{j}}\Bigr|_{\phi=\vev{\phi}, S=\vev{S}} 
=
\begin{pmatrix}
-\frac{m_{0}^4}{8 \pi^2 \vev{\phi}^2}-\lambda_{12} \vev{S}^2
&
\lambda_{12}\vev{S} \vev{\phi} 
\\
\lambda_{12}\vev{S} \vev{\phi}
&
\frac{3 g_{Z'}^4 \vev{S}^2}{8\pi^2}-\lambda_{12} \vev{\phi}^2
\end{pmatrix},
\end{equation}
where we have introduced the abbreviation
\begin{equation}
m_0^4=12 m_t^4-6m_{W}^4- 3 m_Z^4 = (319 \text{ GeV})^{4}.
\end{equation}
Since this matrix is non-diagonal, the mass eigenstates will consist of mixtures of the interaction eigenstates. It is of course simple to write down analytic expressions for the mass eigenvalues
\begin{align}
m_{\pm}^2(\lambda_{12}, g_{Z'}, \vev S) = \frac{1}{2} \text{tr }M^2 \pm \sqrt{\left(\frac{\text{tr }M^2}{2}\right)^2 - \det M^2}, \label{eq:masseigenQP}
\end{align}
but it is not possible to invert them in a closed form for any of the three free parameters. In order to match one of them to the Higgs mass, we will therefore numerically solve the resulting equation.

\begin{figure}
\centering
\includegraphicsbox[width=0.4\textwidth]{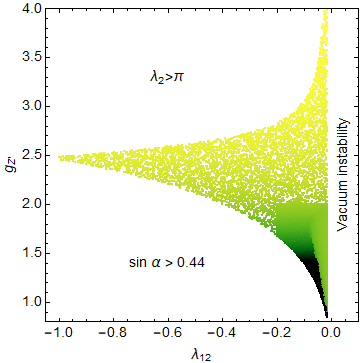}
\includegraphicsbox[width=0.4\textwidth]{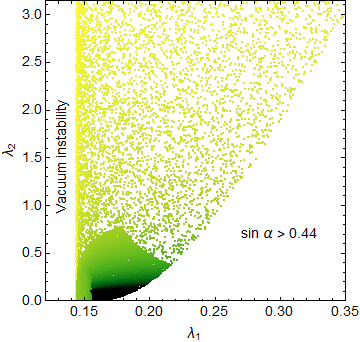}

\vspace{0.3cm}
\includegraphicsbox[width=0.4\textwidth]{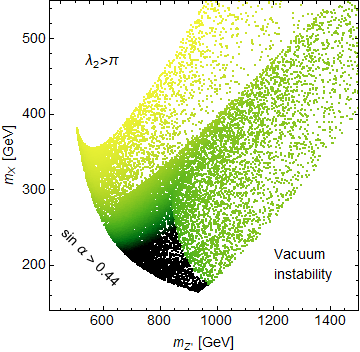}\quad
\includegraphicsbox[width=0.10\textwidth]{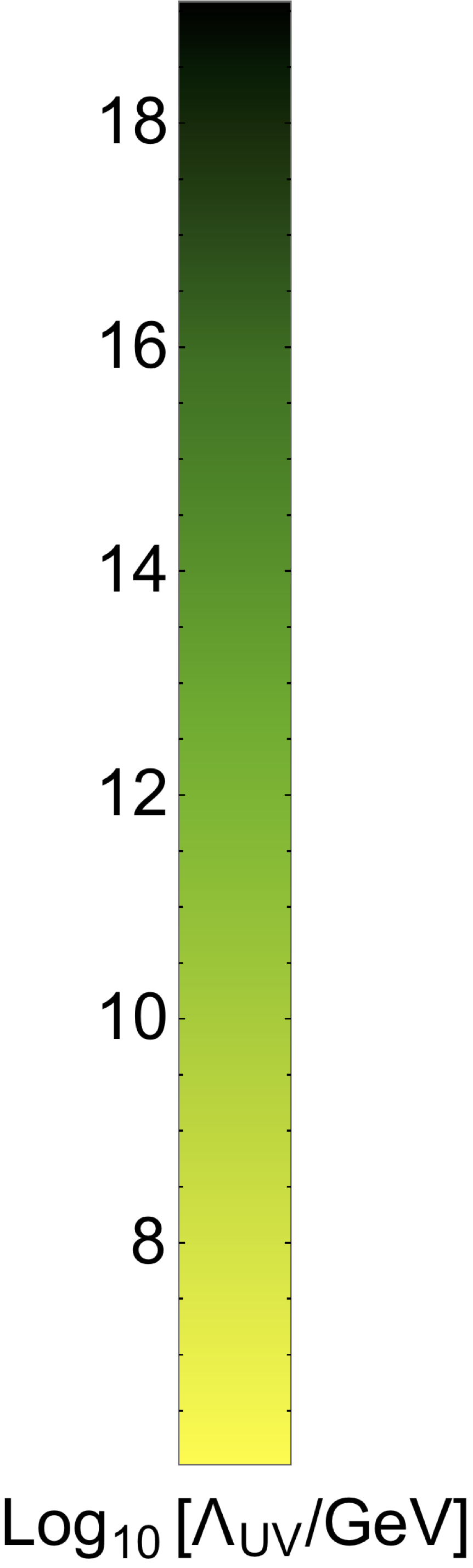}
		\caption{Largest possible UV scales in the QP scenario A of the Hempfling model where
		the Higgs is the lighter scalar particle.}
		\label{fig:QP-Scen-A}
\end{figure}
\begin{figure}[t]
\centering
\includegraphicsbox[width=0.4\textwidth]{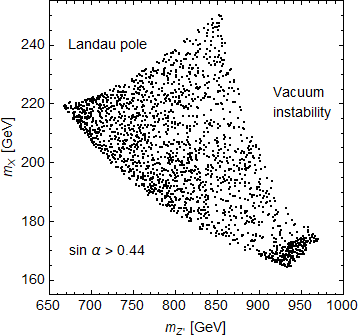}
		\caption{Allowed mass ranges for the new scalar and Z' particles in the Quantum
		Potential scenario A ($m_{X}>m_{h}$) which are perturbatively stable up to the Planck scale. The scenario B  ($m_{X}<m_{h}$) always breaks down before reaching the Planck scale.}
		\label{fig:QP-final}
\end{figure}

\begin{figure}[t]
\centering
			\includegraphicsbox[width=0.4\textwidth]{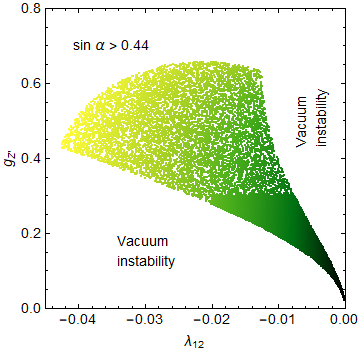}
\includegraphicsbox[width=0.4\textwidth]{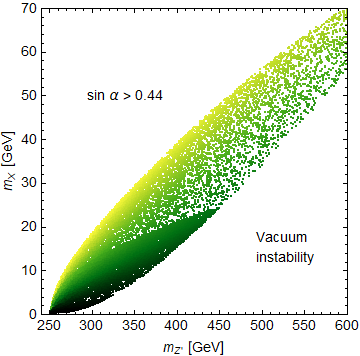}\quad
\includegraphicsbox[width=0.09\textwidth]{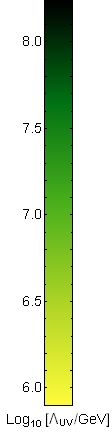}
		\caption{Largest possible UV scales in the QP scenario B of the Hempfling model where
		the Higgs is the heavier scalar particle.}
		\label{fig:QP-Scen-B}
\end{figure}

Now, the presence of two mass eigenvalues opens up two possible scenarios: The Higgs may be lighter or heavier than the new scalar $m_{X}$, i.e.~$m_{h}=m_{\pm}$ and $m_{X}=m_{\mp}$.
We term these scenarios A ($m_{h}=m_{-}$) and B ($m_{h}=m_{+}$). Nevertheless, the numerical procedure is straightforward in both cases: We randomly dial $3\cdot 10^{4}$ pairs  
$\lambda_{12}\in[0,-\pi]$ and $g_{Z'}\in[0,\pi]$ then
\begin{enumerate}
\item Solve the equation
\begin{align}
m_\pm(\lambda_{12}, g_{Z'}, \vev{S}) = m_h
\end{align}
for the expectation value $\vev{S}(\lambda_{12}, g_{Z'})$.
\item Eliminate $\vev S$ in the other mass eigenvalue, resulting in   \begin{align}
m_\mp(\lambda_{12}, g_{Z'}) = m_\mp(\lambda_{12}, g_{Z'}, \vev S(\lambda_{12}, g_{Z'})).
\end{align}
\item For any given pair $\{\lambda_{12}, g_{Z'}\}$ compute the couplings $\{\lambda_1, \lambda_2\}$ from (\ref{eq:statcondQP}) as well as the predicted masses for the new degrees of freedom
\begin{align}
m_X& = m_{\mp}(\lambda_{12}, g_{Z'}), 
& 
m_{Z'} &= g_{Z'} \vev S(\lambda_{12}, g_{Z'}),
\end{align}
which can be used as initial conditions for the RG equations.
\end{enumerate}

By constructing the appropriate parameter regions, we enforce perturbativity
by demanding $|\lambda_i|<\pi$. We also immediately dismiss parameter sets, which lead to unstable extrema of the effective potential (i.e.\ negative values for one of the $m_i^2$) or to sizeable mixing between the Higgs and the new scalar:
\be
m_{\pm}=m_{h}=\cos\alpha \, \phi + \sin \alpha\,  S\, .
\ee
Following the analysis of \cite{Farzinnia:2013pga,Farzinnia:2014xia} this mixing is constrained
to $\sin \alpha < 0.44$ from present experimental bounds.
In addition we study the UV-consistency of the model: We
integrate the beta functions (enforcing the scaling assumption) up to the 
scale at which a Landau pole or vacuum instability occurs and stop at $M_{\text{Planck}}$ if
this does not occur.
This gives the plots shown in \figref{fig:QP-Scen-A} for scenario A and \figref{fig:QP-Scen-B} for scenario B.

In scenario A, where the Higgs is lighter than the new scalar, there is a small  window of allowed couplings which allows for an extrapolation all the way to the Planck scale. This window translates to a range of masses
\begin{align}
650 \text{ GeV} &< m_{Z'} < 970 \text{ GeV}, & 160 \text{ GeV} < m_X < 250 \text{ GeV}\, .
\label{QPfinalmasses}
\end{align}
 for the dark $Z'$-boson and the new scalar resonance, see \figref{fig:QP-final}.
As can be seen from the plots, the scenario B of the new scalar being lighter than the Higgs does not allow for an extrapolation up to the Planck scale at any coupling. 

%%%%%%%%%%%%%%%%%%%%%%%%%%%%%%%%%%%%%%

\section{Gildener--Weinberg Approach}
\label{sec:GW}

Here we study the effective potential of the Hempfling model using the Gildener--Weinberg method of~\cite{Gildener:1976ih}.

%%%%%%%%%%%%%%%%%
\paragraph{Conceptual Idea.}

In their classic quantum field theory paper Gildener and S.~Weinberg 
introduced an elegant formalism to deal with the perturbative construction of the
effective potential in the presence of multiple scalar vacuum expectation values
in classically scale invariant
theories. It represents a generalization of the Coleman--Weinberg idea to the multi-scalar
case and is equivalent to it in the single field case.
The key assumption is that the classical potential 
\be
V_{0}(\vec\Phi)=\frac{1}{4}\lambda_{IJKL}\Phi_{I}\Phi_{J}\Phi_{K}\Phi_{L},
\ee
has a non-trivial minimum at non-zero field values 
$
\vev{\vec{\Phi}}\neq \vec 0
$ at a particular scale $\mu_\text{GW}$, the
Gildener--Weinberg scale. This yields certain relations termed $R$ among the couplings $\lambda_{IJKL}$
\be
\frac{\partial V_{0}}{\partial \Phi_{i}}\Bigr |_{{\mu=\mu_\text{GW}},\vev{\vec\Phi}\neq 0}=0\, 
\quad \Rightarrow \quad
R(\lambda_{IJKL}) \bigr |_{\mu=\mu_\text{GW}} =0\,. 
\ee
Due to scale invariance of $V_{0}(\vec\Phi)$ this immediately implies that one has a vacuum degeneracy of $V_{0}$ along a ray going through the orgin in scalar field space
where the minimal value of the classical potential
is zero
\begin{align}
\vev{\vec{\Phi}} &= \varphi \, \vec{n}\, ,
&
V_{0}\big(\vev{\vec{\Phi}}\big)&=0.
\end{align}
Here $\varphi$ parametrizes the sliding scale and we normalize $\vec{n}^{2}=1$.

This vacuum degeneracy of the classical potential may be lifted by quantum fluctuations.
Evaluating the one-loop contribution to the renormalized effective potential \emph{along} the degenerate vacuum
 ray, one has 
\be
V^{(1)}_{\text{eff}}\big(\vec{\Phi}= \varphi \, \vec{n}\big)= A\, \varphi^{4} + B\,\varphi^{4} \, \ln \frac{\varphi^{2}}{\mu_\text{GW}^{2}}\, .
\label{Veff1gw}
\ee
Here the functions $A$ and $B$ take the form
\begin{align}
A&= \frac{\hbar}{64\pi^{2}\vev{\varphi}^{4}} \sum_{i}
 n_{i}\, m_{i}[\vev{\varphi}\vec{n}]^{4}
\left (  \ln \frac{m_{i}[\vev{\varphi}\vec{n}]^{2}}{\vev{\varphi}^{2}} - a_{i}
\right )
= \frac{\hbar}{64\pi^{2}} \sum_{i}
 n_{i}\, \tilde{m}_{i}[\vec{n}]^{4}
\left (  \ln \tilde{m}_{i}[\vec{n}]^{2}- a_{i}
\right ), \nonumber \\
B&= \frac{\hbar}{64\pi^{2}\vev{\varphi}^{4}} \sum_{i}
 n_{i}\, m_{i}[\vev{\varphi}\vec{n}]^{4}
 = \frac{\hbar}{64\pi^{2}} \sum_{i}
 n_{i}\, \tilde{m}_{i}[\vec{n}]^{4}.
\label{ABdef}
\end{align}
The vacuum expectation value  $\vev{\varphi}$ for the sliding scale field is radiatively generated 
and we generically have $m_{i}[\vev{\varphi}\vec{n}]= \vev{\varphi}\tilde{m}_{i}[\vec{n}]$
such that $A$ and $B$ are in fact independent of $\vev{\varphi}$ and are pure functions of
the couplings.
The extremum of the one-loop effective potential along the ray then lies at
\be
\frac{\vev{\varphi}}{\mu_\text{GW}} = \exp\left [ - \frac{1}{4} - \frac{A}{2B}\right ]\, .
\label{vevvarphidef}
\ee
Hence, as long as $A$ and $B$ are of the same order of magnitude, the logarithm $\ln \frac{\vev{\varphi}}{\mu_\text{GW}}$ in the effective potential stays small and the perturbative expansion is under control.
One also straightforwardly extracts the mass of the excitation along the flat direction $\vec{n}$ which is the pseudo-Goldstone boson of broken scale invariance. Originally massless, its mass
is spontaneously generated by quantum fluctuations and given by the compact expression
\be
m^{2}_\text{PGB}= \frac{\mathrm{d}^{2}V^{(1)}_{\text{eff}}(\varphi \, \vec{n})}{\mathrm{d}\varphi^{2}} \Bigr |_{\varphi=\vev{\varphi}} = 8\, B \, \vev{\varphi}^{2}\, ,
\label{mPGBdef}
\ee
at one-loop precision. Clearly, a positive $B$ is required in order to have a minimum of the
potential.
In conformal extensions of the standard model, $m_\text{PGB}$ may or may not
be identified with the Higgs-mass.

%%%%%%%%%%%%%%%%%%%%%%%%%%

%%%%%%%%%%%%%%%%%
\paragraph{GW for Hempfling Model.}

Let us now analyse the Hempfling model
 in the Gildener--Weinberg approach. The classical
potential for the two real scalars $\phi$ and $S$ reads
\be
V_{0}(\vec{\Phi}) = \frac{\lambda_{1}}{4}\, \phi^{4} 
+\frac{\lambda_{12}}{4}\, \phi^{2}\, S^{2 }+ \frac{\lambda_{2}}{4}\, S^{4} , \qquad
(\lambda_{1}, \lambda_{2} >0)\, ,
\ee
where the positivity constraint on the scalar couplings implies stability.
A degenerate non-trivial vacuum occurs if the condition 
\be
\lambda_{12} = -2 \sqrt{\lambda_{1}\lambda_{2}}\, \label{GWcond1}
\ee
is met at $\mu=\mu_\text{GW}$. At this scale the classical potential takes the simple 
perfect square form 
$$
V_{0}=\bigg(\frac{\sqrt{\lambda_{1}}}{2}\phi^{2} - \frac{\sqrt{\lambda_{2}}}{2}S^{2}\bigg)^{2}\, .
$$
Clearly we then have a degenerate vacuum along the ray

\be
\begin{pmatrix} \phi \cr S \end{pmatrix}_{\text{ray}}
= \varphi\,  \vec{n} =\varphi 
\begin{pmatrix} \cos\alpha \cr \sin\alpha \end{pmatrix}
\, 
=
\frac{\varphi}{\sqrt{\lambda_{1}^{1/2} + \lambda_{2}^{1/2}}}\, 
\begin{pmatrix} \lambda^{1/4}_{2} \cr \lambda^{1/4}_{1} \end{pmatrix}
\, 
 , \quad  \text{with} \quad \tan \alpha= \left (\frac{\lambda_{1}}{\lambda_{2}}\right )^{1/4}
 \, .
\label{GWcond2}
\ee
The particular form of  the one-loop functions $A$ and $B$ in (\ref{Veff1gw}) may be straightforwardly read off from the results in
(\ref{54}) and (\ref{55}). Remarkably one finds that 
\emph{all} the gauge parameter dependent masses $m_{F\pm}$, $m_{G\pm}$ and $m_{I\pm}$ 
vanish identically on the ray (\ref{GWcond2}) upon imposing the relation (\ref{GWcond1})!\footnote{In a related model the gauge invariance in the GW approach was also noted in the appendix of \cite{AlexanderNunneley:2010nw}.}
Hence, in the GW setup with  scaling assumptions 
$\lambda_{s}\sim\mathcal{O}(g^{2}_{i})\sim\mathcal{O}(y^{2}_{t})$ we do find explicit
gauge invariance at leading order in perturbation theory. In addition $m_{E-}$ vanishes as it corresponds
to the tree-level mass of the pseudo-Goldstone boson excitation along the vacuum ray.
For the non-vanishing dimensionless mass coefficients $\tilde m_{i}$ one then finds (enforcing the condition (\ref{GWcond1}) and (\ref{GWcond2}))
\begin{align}
\tilde{m}_A^2 &= \frac{g_2^2 \sqrt{\lambda_{2}}}{4(\sqrt{\lambda_{1}}+\sqrt{\lambda_{2}})} , 
&
\tilde{m}_B^2 &= \frac{\left(g_1^2+g_2^2\right) \sqrt{\lambda_{2}}}{4(\sqrt{\lambda_{1}}+\sqrt{\lambda_{2}})}, 
&
 \tilde{m}_C^2 =  \frac{g_{Z'}^2\sqrt{\lambda_{1}}}{(\sqrt{\lambda_{1}}+\sqrt{\lambda_{2}})},
 \notag \\
\tilde{m}_{E+}^2 &= 2\sqrt{\lambda_{1}\lambda_{2}} , 
&
\tilde{m}_T^2 &= \frac{y_t^2\sqrt{\lambda_{2}}}{2(\sqrt{\lambda_{1}}+\sqrt{\lambda_{2}}) }  .
\end{align}
These are to be inserted into the definitions of the functions $A$ and $B$ in \eqref{ABdef}.
The corresponding masses $m_{i}$ of the $W$ and $Z$ boson, the dark $Z'$ boson, the
scalar $\varphi_{E}$ as well as the top quark $t$ are then obtained by multiplying these
expressions by the vacuum expectation value of the sliding scale $\vev{\varphi}$, to wit
\be
m_{W}=\tilde m_{A}\,\vev{\varphi}\, , \quad
m_{Z}=\tilde m_{B}\,\vev{\varphi}\, , \quad
m_{t}=\tilde m_{T}\,\vev{\varphi}\, , \quad
m_{Z'}=\tilde m_{C}\,\vev{\varphi}\, , \quad
m_{E}=\tilde m_{E+}\,\vev{\varphi}\, .
\ee
 The mass of the pseudo
Goldstone boson then follows from the above and \eqref{mPGBdef}  to be
\be
\label{hier}
m_\text{PGB}^{2}=\frac{6 m_{W}^{4} + 3 m_{Z}^{4} - 12 m_{t}^{4}
+ 3 m_{Z'}^{4} 
+ m_{E}^{4} }{8\pi^{2}\, \vev{\varphi}^{2}}\, .
\ee
 The vacuum expectation value $\vev{\varphi}$ 
is related to the vacuum expectation value of the SM-Higgs field $\vev{\phi}=246 \, \text{GeV}$  
via
\be
\vev{\varphi} =  \frac{\sqrt{\lambda_{1}^{1/2}+\lambda_{2}^{1/2}}}{\lambda_{2}^{1/4}} \vev{\phi} \, .
\ee
This relation then determines the masses of the dark $Z'$ boson and second scalar $\varphi_{E}$
as functions of $\lambda_{1}$ and $\lambda_{2}$ to be
\be
m_{Z'}= g_{Z'} \left (\frac{\lambda_{1}}{\lambda_{2}}\right )^{1/4} \, \vev{\phi}\, ,
\qquad
m_{E}=\sqrt{2}\,  \lambda_{1}^{1/4}\,\sqrt{\lambda_{1}^{1/2}+\lambda_{2}^{1/2}} \, \vev{\phi}\, .
\ee
Moreover, the relation \eqref{vevvarphidef} determines the Glildener-Weinberg
scale $\mu_\text{GW}$ as a function of $\lambda_{1}$ and $\lambda_{2}$ and the
SM parameters:
\be
\mu_\text{GW}=\exp\left [\frac{1}{4}+\frac{A}{2B}\right ]\,  \frac{\sqrt{\lambda_{1}^{1/2}+\lambda_{2}^{1/2}}}{\lambda_{2}^{1/4}} \vev{\phi} .
\label{2.44}
\ee
All unknown quanitities have now been expressed as functions of $\lambda_{1}$ and $\lambda_{2}$. Note that all couplings here are defined at the scale $\mu_\text{GW}$. This means that 
the SM quantities need in principle to be RG evolved from the
electroweak scale  to $\mu_\text{EW}$. However, as long as the relative factor in (\ref{2.44}) is
not too different from 1 this effect may be neglected.

Let us now look at the classical mass matrix in detail. One easily computes
\be
M_{ij}^{2}=  \frac{\partial^{2} V_{0}}{\partial \phi_{i}\partial \phi_{j}}\Bigr|_{\phi=\vev{\phi}, S=\vev{S}} 
=\frac{2\vev{\varphi}^{2}}{\sqrt{\lambda_{1}}+\sqrt{\lambda_{2}}}
\begin{pmatrix}
\lambda_{1}\sqrt{\lambda_{2}} & -(\lambda_{1}\lambda_{2})^{3/4} \cr
 -(\lambda_{1}\lambda_{2})^{3/4} & \lambda_{2}\sqrt{\lambda_{1}}\cr
\end{pmatrix}.
\ee

The two eigenstates corresponding to the scalar masses $m_{E}$ and $m_{\text{PGB}}$ are
expressed in terms of the initial scalar fields $\phi$ and $S$ as
\begin{align}
\label{mixings}
\varphi_{\text{PGB}}= \cos\alpha\, \phi
+\sin\alpha\, S\, ,  \qquad
\varphi_{E}&= -\sin\alpha \, \phi
+ \cos\alpha\, S \, ,\qquad \tan\alpha=\left (\frac{\lambda_{1}}{\lambda_{2}}\right)^{1/4}\, .
\end{align}
Importantly, there are now two options to identify the Higgs mass of $m_{h}=125 \, \text{GeV}$ with the scalar resonances found:  Either  $m_{h}=m_\text{PGB}$ (scenario A) or 
$m_{h}=m_{E}$ (scenario B).
Either choice determines yet another coupling such that in the end all quantities depend
on just two parameters. As discussed in the previous section present observational bounds at the LHC restrict the allowed mixings in the extended Higgs sector. The analysis of 
\cite{Farzinnia:2013pga,Farzinnia:2014xia} restricted the
mixing angle $\omega$ in the parametrization (translated to our conventions) 
\be
h = \cos \omega \, \phi - \sin\omega \, S 
\ee 
to $|\sin\omega| < 0.44$ where $h$ ist the Higgs field mass eigenstate. This 
translates to the mixing angle bounds $|\sin \alpha|<0.44$ in scenario A
and $|\sin \alpha|>0.90$ in scenario B in (\ref{mixings}).

\paragraph{Phenomenological Analysis.}
The key equations are
\begin{align}
3 m_{Z'}^{4}+m_{E}^{4}&= 8\pi^{2}\, m^{2}_{\text{PGB}} \, \frac{\sqrt{\lambda_{1}} + \sqrt{\lambda_{2}}}
{\sqrt{\lambda_{2}}}\, \vev{\phi}^{2}+m_{0}^{4}  \label{(1)} \\
m_{Z'}&= g_{Z'}\, \left (\frac{\lambda_{1}}{\lambda_{2}}\right)^{1/4} \label{(2)} \vev{\phi} \\
m_{E}&= \sqrt{2}\lambda_{1}^{1/4}\, (\sqrt{\lambda_{1}}+\sqrt{\lambda_{2}})^{1/2}\, \vev{\phi}\, , \label{(3)}
\end{align}
where $m_{0}^{4}=12 m_{t}^{4}-6m_{W}^{2}- 3 m_{Z}^{4}= (319 \text{ GeV})^{4}$ and $\vev{\phi}=246 
\text{ GeV}$. It is then clear that the mass of the $Z'$-boson and the mass of the non-Higgs
scalar will depend on a two-parameter family. In scenario A we will take $\{\lambda_{1},\lambda_{2}\}$ while
in scenario B we take $\{\lambda_{1},g_{Z'}\}$ as independent quantities. We randomly generate
values for these couplings and check their perurbative validity by demanding the following
bounds
\be
\vev{\varphi}, \mu_{\text{GW}}\in [24.6 \text{ GeV} , 2460 \text{ GeV}],
\ee
i.e.~these scales are only a factor of ten away from the electroweak scale. Beyond this
we would have to RG evolve the SM parameters to $\mu_\text{GW}$ consistently, which we
did not implement in this work. Moreover,
$\vev{\varphi}$ and $\mu_{\text{GW}}$ are also allowed to differ by a factor of 10 in order to avoid large logarithms. Finally,
we constrain the couplings $|\lambda_{i}|$ and $|g_{Z'}|$ to be numerically smaller than $\pi$.

\begin{figure}[t]
\centering
\includegraphicsbox[width=0.4\textwidth]{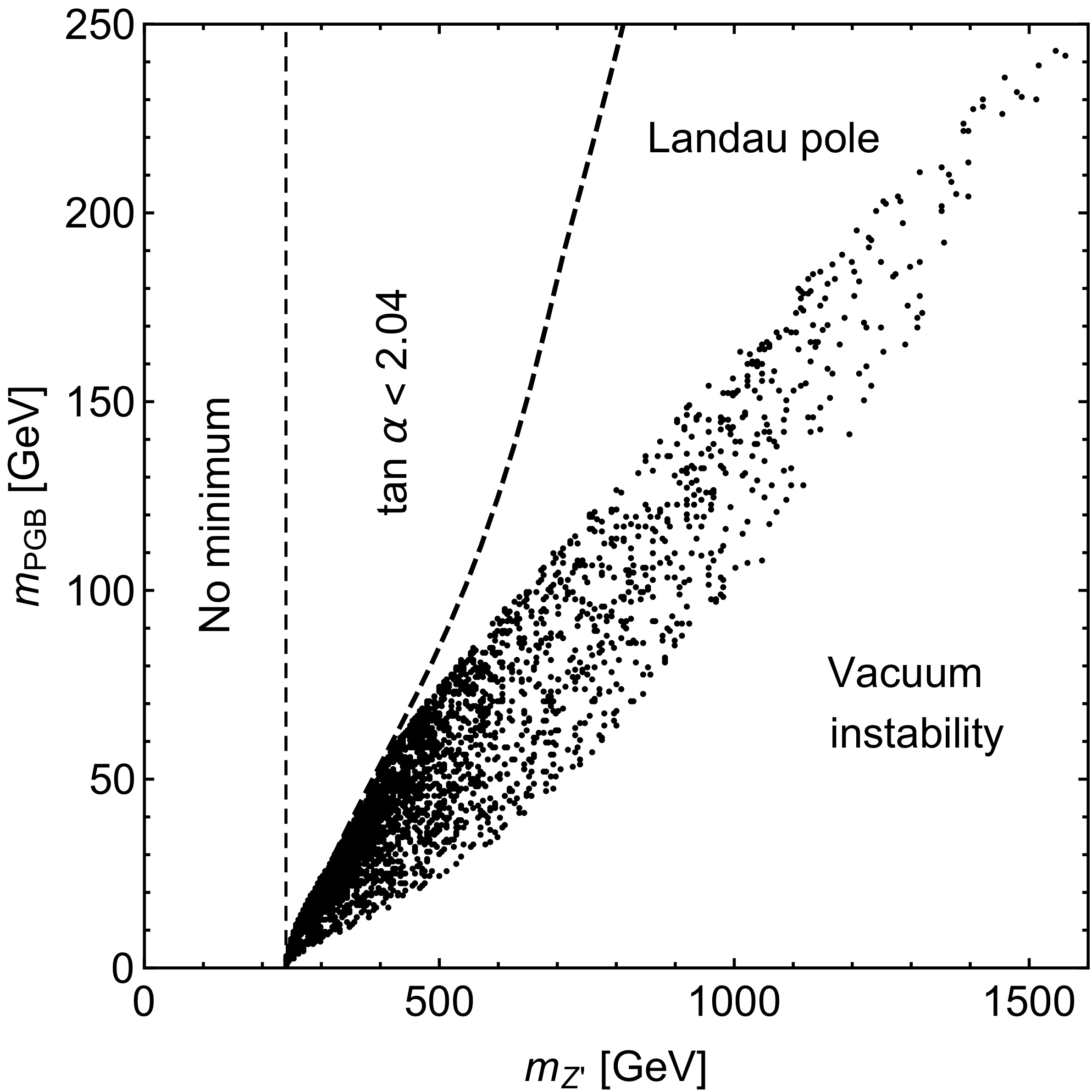}
		\caption{Allowed mass ranges for the new scalar and $Z'$ particles in the Gildener--Weinberg scenario B which are perturbatively stable up to the Planck scale.}
		\label{fig:rugby}
\end{figure}

For every pair $\{\lambda_{1},\lambda_{2}\}$ respectively $\{\lambda_{1},g_{Z'}\}$  
the UV-breakdown scale is computed by integrating the RG equations using the initial
conditions spelled out in the appendix. A breakdown is quantified by any gauge 
coupling becoming larger than 10 (Landau pole) or the scalar couplings $\lambda_{1}$ or
$\lambda_{2}$ turning negative (vacuum instability). For this the one loop RG equations
of the Hempfling model were solved numerically and the breakdown scale $\Lambda_\text{UV}$
recorded for every data point. We analyzed $\mathcal{O}(10^{4})$ random points in both
scenarios:

\begin{itemize}
\item{Scenario A: $m_{h}=m_\text{PGB}$ }

Here we dial a pair $\{\lambda_{1},\lambda_{2}\}$ of couplings to find $m_{E}$ from (\ref{(3)}).
Inserting this into (\ref{(1)}) yields $m_{Z'}$ and using this in (\ref{(2)}) finally gives us 
$g_{Z'}$. The values of  $\{\lambda_{1},\lambda_{2}\}$ were picked randomly in the
interval $[0,\pi]^{2}$. 
However, it turns out that the mixing condition $|\sin\alpha<0.44|$ 
is violated for \emph{all} perturbatively viable resulting pairs $\{\lambda_{1},\lambda_{2}\}$ in this
scenario. Hence this model is ruled out by experiment.

\begin{figure}[t]
\centering
\includegraphicsbox[width=0.4\textwidth]{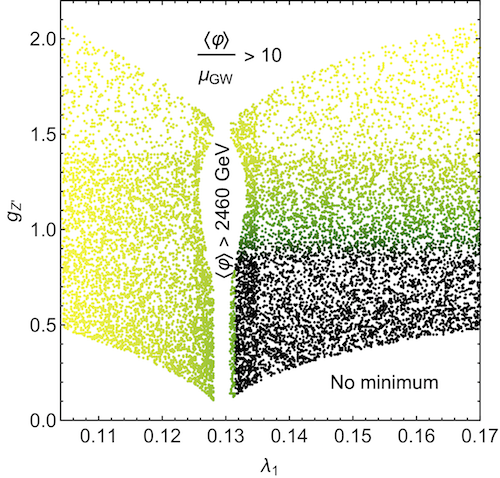}\quad
\includegraphicsbox[width=0.4\textwidth]{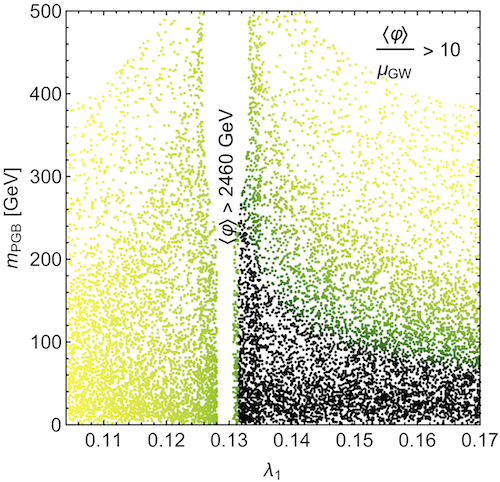}
\includegraphicsbox[width=0.4\textwidth]{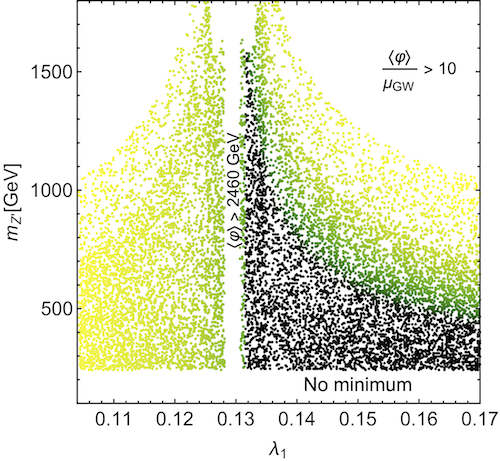}\quad
\includegraphicsbox[width=0.09\textwidth]{Graphics/GW-Scenario-B-scalebar.png}
\caption{Largest possible UV scales in the GW scenario B of the Hempfling model where the Higgs is not the PGB particle. A random set of 12000 dials of $(\lambda_{1},g_{Z'})$ in the realm $\lambda_{1}\in[0.1,0.17]$ and $g_{Z'}\in[0,\pi]$ were performed, tested for perturbative
viability and the UV breakdown scale of every point computed. The non-smooth jump at 
$\lambda_{1}\sim 0.13$ from a UV-cutoff at the Planck scale to around $10^{12}$ GeV is due to
the onset of vacuum instability (negative $\lambda_{1}$ or $\lambda_{2}$) at that 
intermediate scale.
}
		\label{fig:GW-Scen-B}
\end{figure}

\item{Scenario B: $m_{h}=m_E$ }

Now $\lambda_{1}$ and $\lambda_{2}$ are not independent. We therefore dial a pair
$\{g_{Z'},\lambda_{1}\}$ within  $[0,\pi]^{2}$. The coupling $\lambda_{2}$ then follows from
solving (\ref{(2)}) to be given by
\be
\lambda_{2}= \left ( \frac{m_{h}^{2}}{2\sqrt{\lambda_{1}}\, \vev{\phi}^{2}} -
\sqrt{\lambda_{1}} \right )^{2} \, .
\ee
Implementing the mixing constraint $\tan\alpha>2.04$ which amounts to $\lambda_{2}< \lambda_{1}/2.04^{4}$ yields a very narrow range for $\lambda_{1}$:
\be
\lambda_{1}\in [0.104, 0.170].
\ee
This forces $\lambda_{2}$ to be very small, $\lambda_{2}<0.01$ and accordingly 
$|\lambda_{12}|<0.08$.
Then the mass of the $Z'$ boson follows directly from (\ref{(2)}) and the mass of the
non-Higgs scalar $m_{\text{PGB}}$ is deduced from (\ref{(1)}). After checking the RG
evolution of all couplings we plot the breakdown
scale in heat plots of \figref{fig:GW-Scen-B}. One sees that also the gauge coupling
is narrowed down by the UV conditions to $g_{Z'}< 0.9$. Above this value it develops
a Landau pole before $M_{\text{Pl}}$ is reached. 
 Note that the SM value $\lambda_{1}=m^{2}_{h}/2\vev{\phi}^{2}=0.129$
leads to a vanishing $\lambda_{2}$ which entails a diverging $m_{Z'}$ as well as $\mu_{\text{GW}}$. This leads to a departure from the perturbative domain of the
Hempfling model and explains the excluded central regions in the plots
in \figref{fig:GW-Scen-B}. The observed minimum $Z'$-mass follows immediately from (\ref{(1)})
for a vanishing $m_{\text{PGB}}$ to be 
\be
m^{\text{min}}_{Z'}=\sqrt[4]{(12 m_{t}^{4}-6m_{W}^{2}- 3 m_{Z}^{4}-m_{h}^{4})/3}=240.95 
\text{ GeV}\, ,
\ee
which is reproduced in the data. Below this value we have a negative $m_{\text{PGB}}$
and hence no second minimum.

In summary we conclude that the Hempling model in the GW scenario B gives rise to
a perturbatively stable conformal extension of the SM all the way up to the Planck scale.
The allowed values of $m_{Z'}$ and the new scalar resonance $m_{\text{PGB}}$ take a lense
shape with the range
\begin{align}
240 \text{ GeV} &< m_{Z'} < 1600 \text{ GeV}, & 0 \text{ GeV} < m_X < 250 \text{ GeV}\, ,
\end{align}
see \Figref{fig:rugby}. These values are to be contrasted to the Quantum Potential result in (\ref{QPfinalmasses}).

\end{itemize}

\section{Special Case: Single Scalar SM Extension ($g_{Z'}=0$)}
\label{sec:specialcase}

\begin{figure}[t]
	\centering
		\begin{subfigure}[c]{0.4\textwidth}
			\includegraphicsbox[width=1\textwidth]{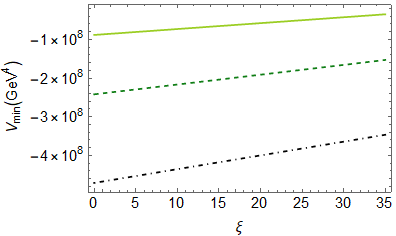}
			\subcaption{Minimum of the effective potential}
		\end{subfigure}
		\hfill
		\begin{subfigure}[c]{0.59\textwidth}
			\includegraphicsbox[width=1\textwidth]{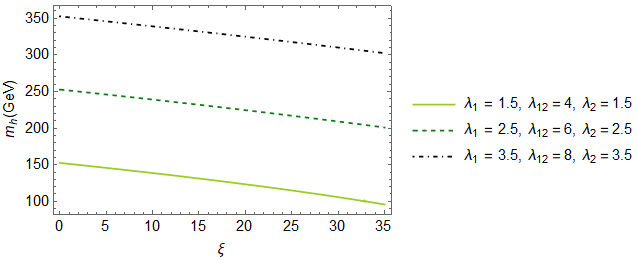}
			\subcaption{Effective mass of the scalar field}
					\end{subfigure}
		\caption{Dependence of the minimum value of the effective potential and effective Higgs mass on the gauge parameter $\xi_B=\xi_W=\xi_{Z'}=\xi$ in the standard model extended by a single scalar field without enforcing a certain hierarchy of couplings.}
		\label{fig:CSM2}
\end{figure}

If we set $g_{Z'}=0$, i.e.~we decouple the dark $Z'$, we arrive at a single scalar extension of the SM Lagrangian. This model was studied in \cite{Meissner:2006zh} and a set of couplings was reported which lead to a minimum of the one-loop effective potential resembling the standard model vacuum. This calculation minimized the sum of the tree and one-loop effective 
potential without enforcing a hierarchy of couplings.
This is problematic as the absence of the QP hierarchy assumption leads to gauge dependence 
of physical data extracted from the effective potential, see fig.~\ref{fig:CSM2}.
The explicit variation of the mass and the minimum value of the potential in the
setup of \cite{Meissner:2006zh} with the gauge paramter $\xi$ is manifest.

However, using the results of the previous sections, it can be seen that independently of the choice of hierarchy, the single scalar conformal standard model without a gauge field does not allow for a stable vacuum consistent with experimental bounds.

First, look at the QP hierachy $\lambda_i \sim g_j^4$. In the limit of vanishing $g_{Z'}$ the mass matrix (\ref{M2QP}) for the scalar field becomes
\begin{align}
M^2_{ij}
=
\begin{pmatrix}
-\frac{m_{0}^4}{8 \pi^2 \vev{\phi}^2}-\lambda_{12} \vev{S}^2
&
\lambda_{12}\vev{S} \vev{\phi}
\\
\lambda_{12}\vev{S} \vev{\phi}
&
-\lambda_{12} \vev{\phi}^2
\end{pmatrix}.
\end{align}
Both mass eigenvalues are real if and only if both the trace and the determinant of this matrix are non-negative. This leads to
\begin{align}
0\leq\frac{\lambda_{12} m_0^4}{8\pi^2}, \qquad\qquad
0\leq- \frac{m_0^4}{8\pi^2 \vev \phi^2} - \lambda_{12} \left(\vev S^2 +\vev \phi^2\right).
\end{align}
Recall that $m_0^4 = (319 \text{ GeV})^4$ is a positive number. Therefore, the first condition can only be met if $\lambda_{12} \geq 0$. But then the second condition is violated, leading us to conclude that within this hierarchy there is no choice of couplings for which both mass eigenvalues turn out positive.

Now we turn to the GW scenario with $\lambda_i \sim g_j^2$. For this assumption we discussed two possible cases: the PGB of scale invariance could either be identified with the Higgs or with the new scalar. 

For the first case, there are no regions of the perturbative parameter space in which the mixing constraint 
\begin{align}
\sin \alpha \leq 0.44
\end{align}
for the mixing angle between $\phi$ and $S$ can be satisfied. This argument is independent of the value of $g_{Z'}$ and stays valid in the decoupling limit.

In the second case, on the other hand, the formula for the mass of the new scalar 
(\ref{hier}) with vanishing $m_{Z'}$ reads
\begin{align}
m_X^2 = \frac{6 m_W^4 + m_{Z}^4 - 12 m_t^4 + m_h^4}{8 \pi^2 \vev \varphi^2} < 0
\end{align}
and predicts a negative mass squared, implying that there is no stable minimum of the effective potential. Hence this model is ruled out.

%%%%%%%%%%%%%%%%%%%%%%%%%

\section{Conclusions}

Since the seminal paper by Coleman and Weinberg \cite{Coleman}, it has been an attractive theoretical concept that mass scales are generated via quantum corrections to a classically scalefree model. Even more appealing, such a mechanism could be part of the theoretical description underlying the nature of electroweak symmetry breaking. In fact, the study of classically conformal modifications of the standard model has recently been subject to great interest. The lack of novel experimental insights puts the identification of a minimal version of such a conformal extension into the spotlight. Though sensitive to the details of an individual definition, this essentially means to minimize the extension of the standard model field content and parameter space, while avoiding conflicts with experimental observations. 

In this paper we have focussed on an assumption that underlies this search for a `minimal conformal standard model', namely the importance of a consistent hierarchy of coupling constants. In particular, we compared two scenarios, each having its own justification: The Gildener--Weinberg method represents the established framework for studying multi-scalar effective potentials and requires the ordinary loop hierarchy of coupling constants $\lambda_s\sim g_i^2$. On the other hand, recent results on gauge invariance in the context of the standard model motivate the alternative hierarchy assumption $\lambda_s\sim g_i^4$, which we dubbed Quantum Potential approach. 

The explicit comparison of the different resulting methodology was performed in the context of the Hempfling model, which represents the historically first phenomenological example of a conformal standard model extension. In addition to the standard model field content, this theory includes a new scalar and a new $\grp{U}(1)$ gauge field. As argued in \Secref{sec:specialcase}, a further reduction of the field content does not result in a phenomenologically viable model. Hence, in this sense the Hempfling model is minimal.
In particular, we explicitly determined and compared the allowed ranges for the new masses and coupling constants, which are compatible with experimental constraints on the scalar mixing angle and the absence of Landau poles and instabilities up to the Planck scale. Notably, both the QP and GW method yield a stable and perturbatively consistent conformal modification of the standard model all the way up to the Planck scale, which reproduces the correct Higgs mass.

\begin{figure}[t]
\centering
\includegraphicsbox[width=0.4\textwidth]{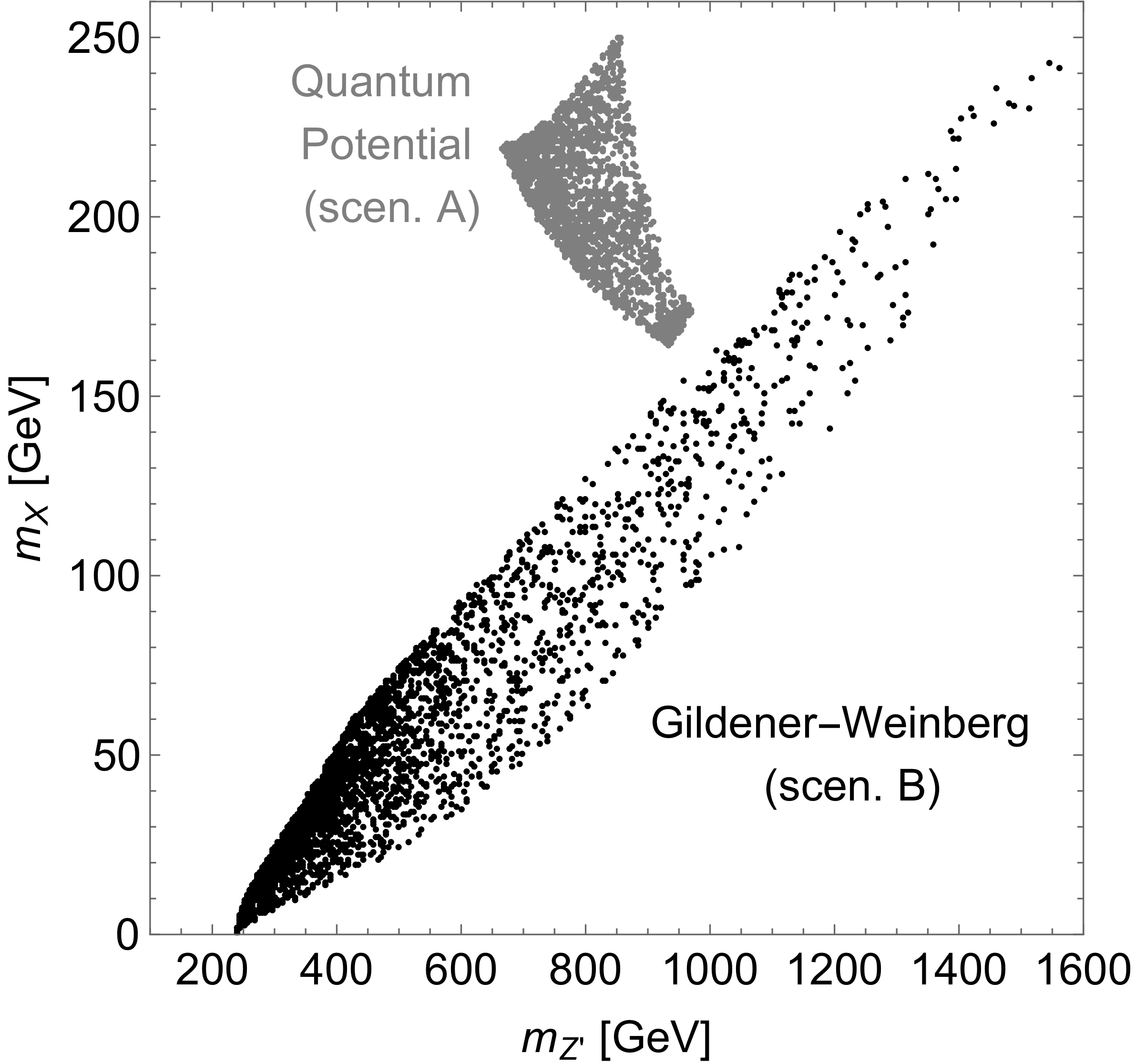}
\caption{Comparison of Gildener--Weinberg (GW) and Quantum Potential (QP) approaches
for the masses of the new scalar and $Z'$ boson in the Hempling model. Only data points with a UV-cutoff at
the Planck scale are plotted. Obviously there is no overlap.
}
\label{fig:GW-QP-compare}
\end{figure}

While arguments in favor of both approaches exist, the obtained results from the GW and QP method show a clear quantitative deviation in the obtained mass constraints and are thus inconsistent with respect to each other, cf.\ \Figref{fig:GW-QP-compare}. Obviously, the allowed combined mass regions for the new particles have no overlap. Still, the allowed intervals for the individual masses of the new gauge field and scalar are not disjoint:
\begin{center}
\includegraphicsbox{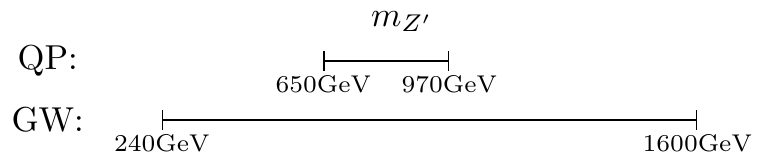}
\qquad
\includegraphicsbox{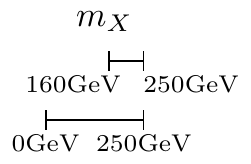}
\end{center}
Hence, the obtained results at least indicate in which mass regions new particles should be expected in order to realize the generation of the electroweak scale by conformal symmetry breaking. 

If different methods yield different phenomenological predictions, what is the correct approach?
Due to the absence of beyond-the-standard-model signals from experiment, the performed analysis does not allow to argue in favor of one of the two approaches using the predicted mass ranges. Rather, the above results stress a generic problem that comes with perturbative methods in general --- and the quest for a minimal conformal standard model in particular --- namely the dependence on a set of assumptions that is hard to prove by analytic means.

We emphasize that as opposed to the much simpler original Coleman--Weinberg model, our hierarchy assumptions have not been proved. Moreover, an analogous reasoning that relies on analytic solutions to the RG-equations seems currently out of reach. It would thus be interesting to explore the consequences of alternative hierarchies, where e.g.\ only one of the scalar couplings $\lambda_i$ is of the order~$g_j^4$. It might also be interesting to draw connections to the functional renormalization group, cf.\ e.g.\ \cite{Gies:2017ajd}.

A very important point is the extension of our analysis to higher loop orders.  While our prescription for the Hempfling model effectively results in a one-scale problem at one-loop order, extending the QP approach to higher loops will most likely face the problem of multi-scale renormalization. This is a clear advantage of the GW method which avoids this problem by construction. On the other hand this fact does not necessarily imply the correctness of the latter approach for generic models. It would thus be important to continue and extend the presented comparison between different methods for the extraction of phenomenological data from effective potentials. Notably, various different ways of treating perturbative calculations with multiple scales have been proposed in the past, see e.g.\ \cite{Einhorn:1983fc,Ford:1994dt,Ford:1996hd,Ford:1996yc,Bando:1992wy,Gildener:1976ih,Chataignier:2018aud}. Very recently, a new method was suggested in \cite{Chataignier:2018aud}, which assumes the existence of a renormalization scale $\mu_\text{CPSS}(\phi)$, where all loop corrections to the effective potential vanish. The vacuum is then obtained by minimizing the tree-level potential $V^{(0)}=\lambda_i\big(\mu_\text{CPSS}(\phi)\big) \phi^4$ with running couplings. It would be very interesting to compare all techniques in a two-loop computation in the context of a simple model. This should help to evaluate the methods' compatibility, perturbative consistency and gauge independence.

While qualitatively pointing at a problem in the identification of phenomenological models via perturbative methods, quantitatively our analysis is certainly incomplete with regard to the gravitational interaction. Importantly, gravity introduces the Planck scale whose naive treatment immediately violates conformal symmetry. Still, proposals for conformal extensions of the standard model exist, which include the dynamic generation of the Planck scale, see e.g.\ \cite{Bars:2013yba,Ferreira:2018itt,Shaposhnikov:2018xkv}.
Moreover, coupling the standard model to gravity necessarily induces higher interaction vertices of the scalar fields, even at the one-loop level \cite{Loebbert:2015eea}. These modify the effective potential and should thus be incorporated into an analysis to draw further conclusions on the viability of the considered models.

%%%%%%%%%%%%%%%%%%%%%%%%%%%%%%%%%%%%%%%%%%%%%%%%%%%%%%%%%%%%%%%%%%%%%%%%%%%
\section*{Acknowledgements}

We thank the CERN Theory Group for kind hospitality during the completion of this work.
The work of FL is funded by the Deutsche Forschungsgemeinschaft (DFG, German Research Foundation) -- Projektnummer 363895012. 
%%%%%%%%%%%%%%%%%%%%%%%%%%%%%%%%%%%%%%%%%%%%%%%%%%%%%%%%%%%%%%%%%%%%%%%%%%%

\appendix

%\section{Hempfling Formulas}

%%%%%%%%%%%%%%%%%%
\section{Phenomenological Data}

Let us collect the phenomenological data which we use throughout the paper \cite{Patrignani:2016xqp}. The masses of the gauge bosons are
\begin{align}
m_W &= 80.385 \pm 0.015 \text{ GeV}, & m_Z = 91.1876 \pm 0.0021 \text{ GeV},
\end{align}
the top mass is
\begin{align}
m_t = 173.1 \pm 0.6 \text{ GeV},
\end{align}
and the mass and vacuum expectation value of the Higgs take the values
\begin{align}
m_H &= 125.09 \pm 0.24 \text{ GeV},& \langle\phi \rangle &= 246.21971 \pm 0.00006 \text{ GeV}.
\label{eq:HiggsMass}
\end{align}
The quartic Higgs coupling $\lambda$ is not known from experiment, but in the standard model it can be deduced from the values for $m_H$ and $\langle \phi \rangle$. In extensions of the standard model this relationship will necessarily be modified; in any case $\lambda$ is not an independent input parameter in the models considered. Since in all our extensions it is still only one Higgs-doublet which couples to the gauge bosons, we assume that $\langle\phi \rangle$ as given in \eqref{eq:HiggsMass} stays the correct expectation value for the interaction eigenstate, even in cases where the mass eigenstate of the Higgs boson differs by some mixing with another scalar. 
For the renormalization group evolution we use the initial values \cite{Buttazzo:2013uya}
\begin{align}
g_1[m_t] &= 0.3583, & g_2[m_t] &= 0.64779, & g_3[m_t] &= 1.1666, & y_t[m_t] = 0.9369
\end{align}
at the $m_{t}$ scale.
%%%%%%

%\remarkf{Is this paragraph necessary/useful?:}
%It is not entirely clear if the masses that are calculated from the effective potential should really be equal to the pole masses of the gauge bosons and the Higgs bosons, as given above, or rather running masses evaluated at the scale of the vacuum expectation value. Anyway, our statements will mostly be of qualitative nature and since all couplings and masses run only logarithmically those statement are not too sensitive to this choice. 
%\remarkjp{ Muessen nicht auch die Werte der $g_{i}$ genutzt werden bei der RGE? Was ist mit
%$\lambda$? Inwiefern ist $\vev{\phi}$ messbar und nicht modellabh\"angig?} \remarkjm{I'm reworking this section}

%%%%%%%%%%%%%%%%%%%%%%%%%%%%%%%%%%%%%%%%%%%%%%%%%%%%%%%%%%%%%%%%%%%%%%%%%%%
%%%%%%%%%%%%%%%%%%%%%%%%%%%%%%%%%%%%%%%%%%%%%%%%%%%%%%%%%%%%%%%%%%%%%%%%%%%
\bibliographystyle{nb}
\bibliography{bibliography}

%bibliography generated by nb.bst v1.06 (C) 2003-2011 Niklas Beisert
\begin{thebibliography}{10}
\providecommand{\href}[2]{#2}
\providecommand{\arxivref}[2]{\href{http://arxiv.org/abs/#1}{#2}}
\providecommand{\doiref}[2]{\href{http://dx.doi.org/#1}{#2}}
\providecommand{\nbbstauthor}[1]{#1}
\providecommand{\nbbstjournal}[1]{\textsf{#1}}
\providecommand{\nbbsttitle}[1]{\textit{#1}}
\providecommand{\nbbsturl}[1]{\texttt{#1}}
\providecommand{\nbbsteprint}[1]{\texttt{#1}}
\providecommand{\nbbststyle}{\raggedright\small\parskip0pt}
\nbbststyle

\bibitem{Degrassi:2012ry}
\nbbstauthor{G.~Degrassi, S.~Di~Vita, J.~Elias-Miro, J.~R.~Espinosa,
  G.~F.~Giudice, G.~Isidori and A.~Strumia},
\nbbsttitle{``{Higgs mass and vacuum stability in the Standard Model at
  NNLO}''},
\nbbstjournal{\doiref{10.1007/JHEP08(2012)098}{JHEP~1208,~098~(2012)}}.

\bibitem{Buttazzo:2013uya}
\nbbstauthor{D.~Buttazzo, G.~Degrassi, P.~P.~Giardino, G.~F.~Giudice, F.~Sala,
  A.~Salvio and A.~Strumia},
\nbbsttitle{``{Investigating the near-criticality of the Higgs boson}''},
\nbbstjournal{\doiref{10.1007/JHEP12(2013)089}{JHEP~1312,~089~(2013)}}.

\bibitem{Coleman:1973jx}
\nbbstauthor{S.~R.~Coleman and E.~J.~Weinberg},
\nbbsttitle{``{Radiative Corrections as the Origin of Spontaneous Symmetry
  Breaking}''},
\nbbstjournal{\doiref{10.1103/PhysRevD.7.1888}{Phys.~Rev.~D7,~1888~(1973)}}.
%%CITATION = PHRVA,D7,1888;%%

\bibitem{Sher:1988mj}
\nbbstauthor{M.~Sher},
\nbbsttitle{``{Electroweak Higgs Potentials and Vacuum Stability}''},
\nbbstjournal{\doiref{10.1016/0370-1573(89)90061-6}{Phys.~Rept.~179,~273~(1989)}}.
%%CITATION = PRPLC,179,273;%%

\bibitem{Bardeen:1995kv}
\nbbstauthor{W.~A.~Bardeen},
\nbbsttitle{``{On naturalness in the standard model}''},
in: \nbbsttitle{``{Ontake Summer Institute on Particle Physics Ontake Mountain,
  Japan, August 27-September 2, 1995}''}.
%%CITATION = FERMILAB-CONF-95-391-T;%%

\bibitem{Fatelo:1994qf}
\nbbstauthor{J.~P.~Fatelo, J.~M.~Gerard, T.~Hambye and J.~Weyers},
\nbbsttitle{``{Symmetry breaking induced by top loops}''},
\nbbstjournal{\doiref{10.1103/PhysRevLett.74.492}{Phys.~Rev.~Lett.~74,~492~(1995)}}.
%%CITATION = PRLTA,74,492;%%

\bibitem{Hambye:1995fr}
\nbbstauthor{T.~Hambye},
\nbbsttitle{``{Symmetry breaking induced by top quark loops from a model
  without scalar mass}''},
\nbbstjournal{\doiref{10.1016/0370-2693(95)01570-1}{Phys.~Lett.~B371,~87~(1996)}},
\nbbsteprint{\arxivref{hep-ph/9510266}{hep-ph/9510266}}.
%%CITATION = HEP-PH/9510266;%%

\bibitem{Hempfling:1996ht}
\nbbstauthor{R.~Hempfling},
\nbbsttitle{``{The Next-to-minimal Coleman-Weinberg model}''},
\nbbstjournal{\doiref{10.1016/0370-2693(96)00446-7}{Phys.~Lett.~B379,~153~(1996)}},
\nbbsteprint{\arxivref{hep-ph/9604278}{hep-ph/9604278}}.
%%CITATION = HEP-PH/9604278;%%

\bibitem{Meissner:2006zh}
\nbbstauthor{K.~A.~Meissner and H.~Nicolai},
\nbbsttitle{``{Conformal Symmetry and the Standard Model}''},
\nbbstjournal{Phys.~Lett.~B648,~312~(2007)}.

\bibitem{Chang:2007ki}
\nbbstauthor{W.-F.~Chang, J.~N.~Ng and J.~M.~S.~Wu},
\nbbsttitle{``{Shadow Higgs from a scale-invariant hidden U(1)(s) model}''},
\nbbstjournal{\doiref{10.1103/PhysRevD.75.115016}{Phys.~Rev.~D75,~115016~(2007)}},
\nbbsteprint{\arxivref{hep-ph/0701254}{hep-ph/0701254}}.
%%CITATION = HEP-PH/0701254;%%

\bibitem{Foot:2007as}
\nbbstauthor{R.~Foot, A.~Kobakhidze and R.~R.~Volkas},
\nbbsttitle{``{Electroweak Higgs as a pseudo-Goldstone boson of broken scale
  invariance}''},
\nbbstjournal{\doiref{10.1016/j.physletb.2007.06.084}{Phys.~Lett.~B655,~156~(2007)}},
\nbbsteprint{\arxivref{0704.1165}{arxiv:0704.1165}}.
%%CITATION = ARXIV:0704.1165;%%

\bibitem{Foot:2007iy}
\nbbstauthor{R.~Foot, A.~Kobakhidze, K.~L.~McDonald and R.~R.~Volkas},
\nbbsttitle{``{A Solution to the hierarchy problem from an almost decoupled
  hidden sector within a classically scale invariant theory}''},
\nbbstjournal{\doiref{10.1103/PhysRevD.77.035006}{Phys.~Rev.~D77,~035006~(2008)}},
\nbbsteprint{\arxivref{0709.2750}{arxiv:0709.2750}}.
%%CITATION = ARXIV:0709.2750;%%

\bibitem{Iso:2009ss}
\nbbstauthor{S.~Iso, N.~Okada and Y.~Orikasa},
\nbbsttitle{``{Classically conformal $B^-$ L extended Standard Model}''},
\nbbstjournal{\doiref{10.1016/j.physletb.2009.04.046}{Phys.~Lett.~B676,~81~(2009)}},
\nbbsteprint{\arxivref{0902.4050}{arxiv:0902.4050}}.
%%CITATION = ARXIV:0902.4050;%%

\bibitem{Iso:2009nw}
\nbbstauthor{S.~Iso, N.~Okada and Y.~Orikasa},
\nbbsttitle{``{The minimal B-L model naturally realized at TeV scale}''},
\nbbstjournal{\doiref{10.1103/PhysRevD.80.115007}{Phys.~Rev.~D80,~115007~(2009)}},
\nbbsteprint{\arxivref{0909.0128}{arxiv:0909.0128}}.
%%CITATION = ARXIV:0909.0128;%%

\bibitem{AlexanderNunneley:2010nw}
\nbbstauthor{L.~Alexander-Nunneley and A.~Pilaftsis},
\nbbsttitle{``{The Minimal Scale Invariant Extension of the Standard Model}''},
\nbbstjournal{\doiref{10.1007/JHEP09(2010)021}{JHEP~1009,~021~(2010)}},
\nbbsteprint{\arxivref{1006.5916}{arxiv:1006.5916}}.
%%CITATION = ARXIV:1006.5916;%%

\bibitem{Carone:2013wla}
\nbbstauthor{C.~D.~Carone and R.~Ramos},
\nbbsttitle{``{Classical scale-invariance, the electroweak scale and vector
  dark matter}''},
\nbbstjournal{\doiref{10.1103/PhysRevD.88.055020}{Phys.~Rev.~D88,~055020~(2013)}},
\nbbsteprint{\arxivref{1307.8428}{arxiv:1307.8428}}.
%%CITATION = ARXIV:1307.8428;%%

\bibitem{Englert:2013gz}
\nbbstauthor{C.~Englert, J.~Jaeckel, V.~V.~Khoze and M.~Spannowsky},
\nbbsttitle{``{Emergence of the Electroweak Scale through the Higgs Portal}''},
\nbbstjournal{\doiref{10.1007/JHEP04(2013)060}{JHEP~1304,~060~(2013)}},
\nbbsteprint{\arxivref{1301.4224}{arxiv:1301.4224}}.
%%CITATION = ARXIV:1301.4224;%%

\bibitem{Farzinnia:2013pga}
\nbbstauthor{A.~Farzinnia, H.-J.~He and J.~Ren},
\nbbsttitle{``{Natural Electroweak Symmetry Breaking from Scale Invariant Higgs
  Mechanism}''},
\nbbstjournal{\doiref{10.1016/j.physletb.2013.09.060}{Phys.~Lett.~B727,~141~(2013)}},
\nbbsteprint{\arxivref{1308.0295}{arxiv:1308.0295}}.
%%CITATION = ARXIV:1308.0295;%%

\bibitem{Hambye:2013sna}
\nbbstauthor{T.~Hambye and A.~Strumia},
\nbbsttitle{``{Dynamical generation of the weak and Dark Matter scale}''},
\nbbstjournal{\doiref{10.1103/PhysRevD.88.055022}{Phys.~Rev.~D88,~055022~(2013)}},
\nbbsteprint{\arxivref{1306.2329}{arxiv:1306.2329}}.
%%CITATION = ARXIV:1306.2329;%%

\bibitem{Heikinheimo:2013fta}
\nbbstauthor{M.~Heikinheimo, A.~Racioppi, M.~Raidal, C.~Spethmann and
  K.~Tuominen},
\nbbsttitle{``{Physical Naturalness and Dynamical Breaking of Classical Scale
  Invariance}''},
\nbbstjournal{\doiref{10.1142/S0217732314500771}{Mod.~Phys.~Lett.~A29,~1450077~(2014)}},
\nbbsteprint{\arxivref{1304.7006}{arxiv:1304.7006}}.
%%CITATION = ARXIV:1304.7006;%%

\bibitem{Holthausen:2013ota}
\nbbstauthor{M.~Holthausen, J.~Kubo, K.~S.~Lim and M.~Lindner},
\nbbsttitle{``{Electroweak and Conformal Symmetry Breaking by a Strongly
  Coupled Hidden Sector}''},
\nbbstjournal{\doiref{10.1007/JHEP12(2013)076}{JHEP~1312,~076~(2013)}},
\nbbsteprint{\arxivref{1310.4423}{arxiv:1310.4423}}.
%%CITATION = ARXIV:1310.4423;%%

\bibitem{Hill:2014mqa}
\nbbstauthor{C.~T.~Hill},
\nbbsttitle{``{Is the Higgs Boson Associated with Coleman-Weinberg Dynamical
  Symmetry Breaking?}''},
\nbbstjournal{\doiref{10.1103/PhysRevD.89.073003}{Phys.~Rev.~D89,~073003~(2014)}},
\nbbsteprint{\arxivref{1401.4185}{arxiv:1401.4185}}.
%%CITATION = ARXIV:1401.4185;%%

\bibitem{Karam:2015jta}
\nbbstauthor{A.~Karam and K.~Tamvakis},
\nbbsttitle{``{Dark matter and neutrino masses from a scale-invariant
  multi-Higgs portal}''},
\nbbstjournal{\doiref{10.1103/PhysRevD.92.075010}{Phys.~Rev.~D92,~075010~(2015)}},
\nbbsteprint{\arxivref{1508.03031}{arxiv:1508.03031}}.
%%CITATION = ARXIV:1508.03031;%%

\bibitem{Oda:2015gna}
\nbbstauthor{S.~Oda, N.~Okada and D.-s.~Takahashi},
\nbbsttitle{``{Classically conformal $U(1)^{\prime}$ extended standard model
  and Higgs vacuum stability}''},
\nbbstjournal{\doiref{10.1103/PhysRevD.92.015026}{Phys.~Rev.~D92,~015026~(2015)}},
\nbbsteprint{\arxivref{1504.06291}{arxiv:1504.06291}}.
%%CITATION = ARXIV:1504.06291;%%

\bibitem{Das:2016zue}
\nbbstauthor{A.~Das, S.~Oda, N.~Okada and D.-s.~Takahashi},
\nbbsttitle{``{Classically conformal U(1)? extended standard model, electroweak
  vacuum stability, and LHC Run-2 bounds}''},
\nbbstjournal{\doiref{10.1103/PhysRevD.93.115038}{Phys.~Rev.~D93,~115038~(2016)}},
\nbbsteprint{\arxivref{1605.01157}{arxiv:1605.01157}}.
%%CITATION = ARXIV:1605.01157;%%

\bibitem{Helmboldt:2016mpi}
\nbbstauthor{A.~J.~Helmboldt, P.~Humbert, M.~Lindner and J.~Smirnov},
\nbbsttitle{``{Minimal conformal extensions of the Higgs sector}''},
\nbbstjournal{\doiref{10.1007/JHEP07(2017)113}{JHEP~1707,~113~(2017)}},
\nbbsteprint{\arxivref{1603.03603}{arxiv:1603.03603}}.
%%CITATION = ARXIV:1603.03603;%%

\bibitem{Hambye:2018qjv}
\nbbstauthor{T.~Hambye, A.~Strumia and D.~Teresi},
\nbbsttitle{``{Super-cool Dark Matter}''},
\nbbsteprint{\arxivref{1805.01473}{arxiv:1805.01473}}.
%%CITATION = ARXIV:1805.01473;%%

\bibitem{Antipin:2013exa}
\nbbstauthor{O.~Antipin, M.~Mojaza and F.~Sannino},
\nbbsttitle{``{Conformal Extensions of the Standard Model with Veltman
  Conditions}''},
\nbbstjournal{\doiref{10.1103/PhysRevD.89.085015}{Phys.~Rev.~D89,~085015~(2014)}},
\nbbsteprint{\arxivref{1310.0957}{arxiv:1310.0957}}.
%%CITATION = ARXIV:1310.0957;%%

\bibitem{Chankowski:2014fva}
\nbbstauthor{P.~H.~Chankowski, A.~Lewandowski, K.~A.~Meissner and H.~Nicolai},
\nbbsttitle{``{Softly broken conformal symmetry and the stability of the
  electroweak scale}''},
\nbbstjournal{\doiref{10.1142/S0217732315500066}{Mod.~Phys.~Lett.~A30,~1550006~(2015)}},
\nbbsteprint{\arxivref{1404.0548}{arxiv:1404.0548}}.
%%CITATION = ARXIV:1404.0548;%%

\bibitem{Einhorn:1983fc}
\nbbstauthor{M.~B.~Einhorn and D.~R.~T.~Jones},
\nbbsttitle{``{A New Renormalization Group Approach to Multiscale Problems}''},
\nbbstjournal{\doiref{10.1016/0550-3213(84)90127-5}{Nucl.~Phys.~B230,~261~(1984)}}.
%%CITATION = NUPHA,B230,261;%%

\bibitem{Ford:1994dt}
\nbbstauthor{C.~Ford},
\nbbsttitle{``{Multiscale renormalization group improvement of the effective
  potential}''},
\nbbstjournal{\doiref{10.1103/PhysRevD.50.7531}{Phys.~Rev.~D50,~7531~(1994)}},
\nbbsteprint{\arxivref{hep-th/9404085}{hep-th/9404085}}.
%%CITATION = HEP-TH/9404085;%%

\bibitem{Ford:1996hd}
\nbbstauthor{C.~Ford and C.~Wiesendanger},
\nbbsttitle{``{A Multiscale subtraction scheme and partial renormalization
  group equations in the O(N) symmetric phi**4 theory}''},
\nbbstjournal{\doiref{10.1103/PhysRevD.55.2202}{Phys.~Rev.~D55,~2202~(1997)}},
\nbbsteprint{\arxivref{hep-ph/9604392}{hep-ph/9604392}}.
%%CITATION = HEP-PH/9604392;%%

\bibitem{Ford:1996yc}
\nbbstauthor{C.~Ford and C.~Wiesendanger},
\nbbsttitle{``{Multiscale renormalization}''},
\nbbstjournal{\doiref{10.1016/S0370-2693(97)00237-2}{Phys.~Lett.~B398,~342~(1997)}},
\nbbsteprint{\arxivref{hep-th/9612193}{hep-th/9612193}}.
%%CITATION = HEP-TH/9612193;%%

\bibitem{Casas:1998cf}
\nbbstauthor{J.~A.~Casas, V.~Di~Clemente and M.~Quiros},
\nbbsttitle{``{The Effective potential in the presence of several mass
  scales}''},
\nbbstjournal{\doiref{10.1016/S0550-3213(99)00262-X}{Nucl.~Phys.~B553,~511~(1999)}},
\nbbsteprint{\arxivref{hep-ph/9809275}{hep-ph/9809275}}.
%%CITATION = HEP-PH/9809275;%%

\bibitem{Bando:1992wy}
\nbbstauthor{M.~Bando, T.~Kugo, N.~Maekawa and H.~Nakano},
\nbbsttitle{``{Improving the effective potential: Multimass scale case}''},
\nbbstjournal{\doiref{10.1143/PTP.90.405,
  10.1143/ptp/90.2.405}{Prog.~Theor.~Phys.~90,~405~(1993)}},
\nbbsteprint{\arxivref{hep-ph/9210229}{hep-ph/9210229}}.
%%CITATION = HEP-PH/9210229;%%

\bibitem{Gildener:1976ih}
\nbbstauthor{E.~Gildener and S.~Weinberg},
\nbbsttitle{``{Symmetry Breaking and Scalar Bosons}''},
\nbbstjournal{\doiref{10.1103/PhysRevD.13.3333}{Phys.~Rev.~D13,~3333~(1976)}}.
%%CITATION = PHRVA,D13,3333;%%

\bibitem{Chataignier:2018aud}
\nbbstauthor{L.~Chataignier, T.~Prokopec, M.~G.~Schmidt and B.~Swiezewska},
\nbbsttitle{``{Single-scale Renormalisation Group Improvement of Multi-scale
  Effective Potentials}''},
\nbbsteprint{\arxivref{1801.05258}{arxiv:1801.05258}}.
%%CITATION = ARXIV:1801.05258;%%

\bibitem{Jackiw:1974cv}
\nbbstauthor{R.~Jackiw},
\nbbsttitle{``{Functional evaluation of the effective potential}''},
\nbbstjournal{\doiref{10.1103/PhysRevD.9.1686}{Phys.~Rev.~D9,~1686~(1974)}}.
%%CITATION = PHRVA,D9,1686;%%

\bibitem{Dolan:1974gu}
\nbbstauthor{L.~Dolan and R.~Jackiw},
\nbbsttitle{``{Gauge Invariant Signal for Gauge Symmetry Breaking}''},
\nbbstjournal{\doiref{10.1103/PhysRevD.9.2904}{Phys.~Rev.~D9,~2904~(1974)}},
[,21(1974)].
%%CITATION = PHRVA,D9,2904;%%

\bibitem{Andreassen:2014eha}
\nbbstauthor{A.~Andreassen, W.~Frost and M.~D.~Schwartz},
\nbbsttitle{``{Consistent Use of Effective Potentials}''},
\nbbstjournal{\doiref{10.1103/PhysRevD.91.016009}{Phys.~Rev.~D91,~016009~(2015)}},
\nbbsteprint{\arxivref{1408.0287}{arxiv:1408.0287}}.
%%CITATION = ARXIV:1408.0287;%%

\bibitem{Andreassen:2014gha}
\nbbstauthor{A.~Andreassen, W.~Frost and M.~D.~Schwartz},
\nbbsttitle{``{Consistent Use of the Standard Model Effective Potential}''},
\nbbstjournal{\doiref{10.1103/PhysRevLett.113.241801}{Phys.~Rev.~Lett.~113,~241801~(2014)}},
\nbbsteprint{\arxivref{1408.0292}{arxiv:1408.0292}}.
%%CITATION = ARXIV:1408.0292;%%

\bibitem{Chataignier:2018kay}
\nbbstauthor{L.~Chataignier, T.~Prokopec, M.~G.~Schmidt and B.~Swiezewska},
\nbbsttitle{``{Systematic analysis of radiative symmetry breaking in models
  with extended scalar sector}''},
\nbbsteprint{\arxivref{1805.09292}{arxiv:1805.09292}}.
%%CITATION = ARXIV:1805.09292;%%

\bibitem{Srednicki}
\nbbstauthor{M.~Srednicki},
\nbbsttitle{``{Quantum field theory}''},
Cambridge University Press (2007).

\bibitem{Farzinnia:2014xia}
\nbbstauthor{A.~Farzinnia and J.~Ren},
\nbbsttitle{``{Higgs Partner Searches and Dark Matter Phenomenology in a
  Classically Scale Invariant Higgs Boson Sector}''},
\nbbstjournal{\doiref{10.1103/PhysRevD.90.015019}{Phys.~Rev.~D90,~015019~(2014)}},
\nbbsteprint{\arxivref{1405.0498}{arxiv:1405.0498}}.
%%CITATION = ARXIV:1405.0498;%%

\bibitem{Coleman}
\nbbstauthor{S.~R.~Coleman and E.~J.~Weinberg},
\nbbsttitle{``{Radiative Corrections as the Origin of Spontaneous Symmetry
  Breaking}''},
\nbbstjournal{Phys.~Rev.~D7,~1888~(1973)}.

\bibitem{Gies:2017ajd}
\nbbstauthor{H.~Gies and R.~Sondenheimer},
\nbbsttitle{``{Renormalization Group Flow of the Higgs Potential}''},
\nbbstjournal{\doiref{10.1098/rsta.2017.0120}{Phil.~Trans.~Roy.~Soc.~Lond.~A376,~20170120~(2018)}},
\nbbsteprint{\arxivref{1708.04305}{arxiv:1708.04305}},
in: \nbbsttitle{``{Proceedings, Higgs cosmology: Theo Murphy meeting:
  Buckinghamshire, UK, March 27-28, 2017}''},
pp.~20170120.
%%CITATION = ARXIV:1708.04305;%%

\bibitem{Bars:2013yba}
\nbbstauthor{I.~Bars, P.~Steinhardt and N.~Turok},
\nbbsttitle{``{Local Conformal Symmetry in Physics and Cosmology}''},
\nbbstjournal{\doiref{10.1103/PhysRevD.89.043515}{Phys.~Rev.~D89,~043515~(2014)}},
\nbbsteprint{\arxivref{1307.1848}{arxiv:1307.1848}}.
%%CITATION = ARXIV:1307.1848;%%

\bibitem{Ferreira:2018itt}
\nbbstauthor{P.~G.~Ferreira, C.~T.~Hill and G.~G.~Ross},
\nbbsttitle{``{Inertial Spontaneous Symmetry Breaking and Quantum Scale
  Invariance}''},
\nbbsteprint{\arxivref{1801.07676}{arxiv:1801.07676}}.
%%CITATION = ARXIV:1801.07676;%%

\bibitem{Shaposhnikov:2018xkv}
\nbbstauthor{M.~Shaposhnikov and A.~Shkerin},
\nbbsttitle{``{Conformal symmetry: towards the link between the Fermi and the
  Planck scales}''},
\nbbsteprint{\arxivref{1803.08907}{arxiv:1803.08907}}.
%%CITATION = ARXIV:1803.08907;%%

\bibitem{Loebbert:2015eea}
\nbbstauthor{F.~Loebbert and J.~Plefka},
\nbbsttitle{``{Quantum Gravitational Contributions to the Standard Model
  Effective Potential and Vacuum Stability}''},
\nbbstjournal{\doiref{10.1142/S0217732315501898}{Mod.~Phys.~Lett.~A30,~1550189~(2015)}},
\nbbsteprint{\arxivref{1502.03093}{arxiv:1502.03093}}.
%%CITATION = ARXIV:1502.03093;%%

\bibitem{Patrignani:2016xqp}
Particle Data Group Collaboration, \nbbstauthor{C.~Patrignani et~al.},
\nbbsttitle{``{Review of Particle Physics}''},
\nbbstjournal{\doiref{10.1088/1674-1137/40/10/100001}{Chin.~Phys.~C40,~100001~(2016)}}.
%%CITATION = CHPHD,C40,100001;%%

\end{thebibliography}

\end{document}